\documentclass[aps,twocolumn,prd,superscriptaddress]{revtex4-2}

\usepackage{diagbox}

\usepackage{mathtools}
\usepackage{amsfonts}
\usepackage{mathrsfs}
\usepackage{bbm}
\usepackage{slashed}
\usepackage{dsfont}

\usepackage{graphicx}
\usepackage{color}
\usepackage{array}

\usepackage{blindtext}
\usepackage{placeins}
\usepackage{booktabs}
\usepackage{makecell}
\usepackage{epstopdf}
\usepackage[caption=false]{subfig}
\usepackage{xspace}
\usepackage{siunitx}
\usepackage{hyperref}
\usepackage[nameinlink]{cleveref}
\usepackage{appendix}

\usepackage{xifthen}
\usepackage{xcolor}
\hypersetup{
	colorlinks,
	linkcolor={red!75!black},
	citecolor={blue!75!black},
	urlcolor={blue!75!black}
}

\setkeys{Gin}{width=0.48\textwidth}

\graphicspath{
	{./figures/}
	{../figures/}
}

\def\Drbar{{\bar{D}\!\llap{/}}\,}

\def\s0#1#2{\mbox{\small{$ \frac{#1}{#2} $}}}
\def\0#1#2{\frac{#1}{#2}}

\newcommand{\Tr}{{\text{Tr}}}

\newcommand{\sumint}{\int\!\!\!\!\!\!\!\!\sum}

\newcommand{\imag}{\text{i}}

\usepackage{multirow}
\usepackage{colortbl}
\definecolor{kugray5}{RGB}{224,224,224}
\newcommand{\PreserveBackslash}[1]{\let\temp=\\#1\let\\=\temp}
\newcolumntype{C}[1]{>{\PreserveBackslash\centering}p{#1}}
\newcolumntype{R}[1]{>{\PreserveBackslash\raggedleft}p{#1}}
\newcolumntype{L}[1]{>{\PreserveBackslash\raggedright}p{#1}}

\newcommand{\spatial}[1]{\boldsymbol{#1}}
\newcommand{\tr}{{\text{tr}}}

\begin{document}

\title{Finite density signatures of confining and chiral dynamics in QCD thermodynamics and fluctuations of conserved charges}

\author{Yi Lu}
\email{qwertylou@pku.edu.cn}
\affiliation{Department of Physics and State Key Laboratory of Nuclear Physics and Technology, Peking University, Beijing 100871, China}

\author{Fei Gao}
\email[Corresponding author: ]{fei.gao@bit.edu.cn}
\affiliation{School of Physics, Beijing Institute of Technology, 100081 Beijing, China}

\author{Yu-xin Liu}
\email{yxliu@pku.edu.cn}
\affiliation{Department of Physics and State Key Laboratory of Nuclear Physics and Technology, Peking University, Beijing 100871, China}
\affiliation{Center for High Energy Physics, Peking University, 100871 Beijing, China}
\affiliation{Collaborative Innovation Center of Quantum Matter, Beijing 100871, China}

\author{Jan M. Pawlowski}
\email{J.Pawlowski@thphys.uni-heidelberg.de}
\affiliation{Institut f{\"u}r Theoretische Physik,
	Universit{\"a}t Heidelberg, Philosophenweg 16,
	69120 Heidelberg, Germany
}
\affiliation{ExtreMe Matter Institute EMMI,
	GSI, Planckstr. 1,
	64291 Darmstadt, Germany
}

\begin{abstract}

We evaluate thermodynamic observables such as pressure, baryon number, entropy and energy density, as well as the second and fourth order baryon number cumulants in the phase structure of QCD. The intertwined confinement and chiral dynamics is resolved within functional QCD, aiming for quantitative accuracy at larger densities. Specifically it is shown that the self-consistent resolution of the confining gluonic background is crucial in particular for even the qualitative  properties of the cumulants. Our results are in quantitative agreement with lattice and functional QCD benchmarks at vanishing and small chemical potentials. Moreover, they offer novel insights in the dynamics at larger chemical potentials including the regime of the critical end point. A welcome by-product of this analysis is the computation of the Polyakov loop potential in finite density QCD, which, alongside the aforementioned observables, can be used as input and benchmark for effective theory computations at finite density. 

\end{abstract}

\maketitle

\section{introduction}

With the new heavy ion experiments CBM (FAIR) and CEE+ (HIAF) being nearly completed, experimental heavy ion physics at collision energies 
$\sqrt{s_{NN}}$ between 3 and 5\,GeV is entering an exciting precision physics era. Importantly, this energy regime encompasses baryon chemical potentials of $600 \lesssim \mu_B\lesssim 650$\,MeV, where the critical end point (CEP) or, more precisely, the onset of new QCD phases is expected, see \cite{Fu:2019hdw, Gao:2020fbl, Gunkel:2021oya}.   The data from CBM or CEE+ will provide information on correlation of conserved charges and further observables after the chemical freeze-out. The resolution or rather reconstruction of the underlying strongly correlated QCD dynamics requires an abundance of experimental data that is accompanied with respective theoretical precision results. The latter endeavour asks for first principle QCD computations of thermodynamic observables and fluctuations of conserved charges at collision energies  $3\,\textrm{GeV} \lesssim \sqrt{s_{NN}}\lesssim 5$\,GeV, that roughly translates into baryon chemical potentials of $500 \lesssim \mu_B\lesssim 700$\,MeV. 

Thermodynamic observables such as the pressure, entropy and energy density as well as the baryon charge as well as fluctuations of conserved charges encode much of the confining and chiral dynamics of QCD, see e.g.~\cite{Fu:2016tey, Isserstedt:2019pgx, Braun-Munzinger:2020jbk, Bazavov:2020bjn, Sorensen:2021zme, Vovchenko:2021kxx, Bernhardt:2022mnx,  Gupta:2022phu, Li:2023mpv, Xu:2023vxy, Fu:2023lcm, Borsanyi:2023wno}. Moreover, their equilibrium values serve as QCD input for transport and hydrodynamic computations~\cite{Vovchenko:2021kxx,Wu:2021xgu}. In particular the fluctuations of conserved charges are well accessible in the experiment~\cite{STAR:2022vlo,STAR:2021fge}, and, as mentioned above, we expect a plethora of respective data in the next decade. 
To date, \textit{direct} QCD computations in this regime can only be performed with functional QCD approaches, as lattice simulations still suffer from the sign problem. 
Moreover, by definition, any extrapolation from lower chemical potentials is explicitly or implicitly based on smoothness assumptions. This makes it structurally unattainable to accommodate new physics, leaving aside systematic error estimates.  

The present work concentrates on the first, important part of full analysis of the experimental signatures of the confining and chiral dynamics at finite density. It is concerned with the computation of observables in equilibrium QCD, using the  self-consistent functional QCD approach based on Dyson-Schwinger equations (DSEs)  developed in \cite{Gao:2020qsj, Gao:2020fbl, Gao:2021wun, Lu:2023mkn}. Specific emphasis is given to the computation and evaluation of correlation functions on the quantum equations of motion of QCD, including the confining glue background, see \cite{Braun:2007bx, Fister:2013bh, Fischer:2013eca, Fischer:2014ata, Fischer:2014vxa, Herbst:2015ona}. This setup allows us to also access  observables that depend to a sizeable extend on the confining dynamics or are even dominated by it. In \Cref{sec:PhaseStructure} we present our results for the phase structure with both the chiral and confinement-deconfinement crossovers. These results are obtained on the quantum equations of motion of QCD with a non-trivial temporal gauge field background, whose effective potential, the Polyakov loop potential, is also detailed here. Part of this analysis is also presented in \Cref{app:DSE-Polya}.  In \Cref{sec:ResultsObservables} we discuss our results for the thermodynamic observables and the fluctuations of conserved charges in the phase structure of QCD, including the regime around the CEP. This discussion also includes an in detail analysis of the importance of the confining background for the results. In \Cref{sec:Summary}, we briefly summarise the main results and provide an outlook on the next steps in the programme of uncovering the QCD physics in the CEP regime.

%%%%%%%%%%%%%%%%%%%%%%%%%%%%%%%%%
\section{QCD Correlation functions in the phase structure of QCD}
\label{sec:CorrelationfunctionsPhaseStructure}

In this Chapter we discuss the chiral and confinement-deconfinement phase structure of QCD. The respective results for correlation functions and order parameters are the basis for the computation of thermodynamic observables and the fluctuations of conserved charges in the phase structure of QCD in \Cref{sec:ResultsObservables}. We focus on finite baryon chemical potential $\mu_B$ for vanishing strangeness and charge chemical potentials $\mu_Q=\mu_S=0$, with the quark chemical potentials 
\begin{align}
	\mu_u = \mu_d = \mu_s = \mu_q \,,\qquad \mu_q = \frac{1}{3} \mu_B\,. 
	\label{eq:muqs}
\end{align}
In a follow-up work we will also discuss the physically interesting case of strangeness neutrality, see e.g.~\cite{Rennecke:2019dxt}. 

In \Cref{sec:Fun+Conf} we briefly summarise the functional framework that gives access to the confinement-deconfinement properties of QCD. In \Cref{sec:Conf-Deconf} this approach is put to work for the computation of the confinement-deconfinement phase structure, and in \Cref{sec:PhaseStructureResults} we present our results for the chiral and confinement-deconfinement order parameters and the phase structure of QCD.

%%%%%%%%%%%%%%%%%%%%
\subsection{Correlation functions and confining dynamics in functional approaches}
\label{sec:Fun+Conf}

In functional approaches, the resolution of the phase structure as well as more generally the computation of observables is done in a two-step process: In the first step the gauge-fixed correlation functions of quarks, gluons and possibly that of composite degrees of freedom are computed from field derivatives of the functional relations for the effective action, the DSEs or functional renormalisation group (fRG) flows. In a second step one computes gauge invariant observables from their representation in terms of these correlation functions. While the information of the phase structure is readily extracted even from the gauge-fixed correlation functions themselves, the diagrammatic representations of general observables can be rather convoluted. Importantly, the underlying expansion schemes are optimised, if the correlation functions are computed on the quantum equations of motion (EoM) of QCD: then, observables are expanded about the global minimum of the theory, which ensures the most rapid convergence of these computations. In short, while a computation of the correlation functions in the background of the solution of the quantum EoM is not required, in a given approximation it may even be crucial for obtaining fully converged results instead of a qualitative failure for some observables. Indeed, this is the case for observables that carry confining information or are even dominated by the confining dynamics such as the kurtosis of net-baryon number. 

The property of the optimal expansion point is readily derived from the functional Dyson-Schwinger equation, from which all DSEs for the correlation functions are derived. Structurally it reads 
\begin{align} 
	\frac{\delta\Gamma[\bar A,\Phi]}{\delta \Phi_i}= \left \langle \frac{\delta S_\textrm{QCD}[\bar A,\hat\Phi]}{\delta \hat \Phi_i }\right\rangle \,, 
	\label{eq:FunDSE}
\end{align} 
where $S_\textrm{QCD}[\bar A,\Phi]$ is the classical action of gauge fixed QCD, see \labelcref{eq:ClassicalAction} in \Cref{app:DSE-Polya} and $\Gamma[\bar A,\Phi]$ is the quantum effective action of QCD. In the present work we use a background field approach for accommodating the non-trivial gluonic background: the gauge field is decomposed into a background $\bar A$ and a dynamical fluctuation field $a$ with $A_\mu=\bar A_\mu+a_\mu$, and the $\bar A$ also enters the gauge fixing condition, see 	\labelcref{eq:Backgroundgauge} in \Cref{app:DSE-Polya}. The  fluctuation (mean) superfield $\Phi$ contains all fundamental fields in QCD: gluons, ghosts and quarks,  
\begin{align}
	\Phi=\langle \hat \Phi\rangle =(a_\mu, c,\bar c, q, \bar q)\,, 
	\label{eq:SuperField}
\end{align}
We note that the DSE \labelcref{eq:FunDSE} for the fluctuation field is accompanied by that of the background field, see \labelcref{eq:A0DSE} in \Cref{app:DSE-Polya}. 
The latter DSE is important for the computation of the order parameter potential  of the confinement-deconfinement phase transition, while the former are used for computing the vertices and propagators of the fundamental fields in QCD. 
In summary, the background field approach facilitates the access to the confining dynamics of QCD as discussed in \Cref{app:Poloop+PolPot} and below. 
The quantum equations of motions of QCD are given by 
\begin{align} 
\frac{\delta\Gamma[\bar A,\Phi]}{\delta \Phi_i}= 0 =  \frac{\delta\Gamma[\bar A,\Phi]}{\delta \bar A}\,,
\label{eq:QEoM}
\end{align} 
with the solution $(\bar A,\Phi)=(\bar A,\Phi)_\textrm{EoM}$. \Cref{eq:QEoM} implies that the right hand sides of \labelcref{eq:FunDSE,eq:A0DSE} vanish on the quantum EoM. Put differently, the contributions of the off-shell fluctuations to the effective action vanish on-shell. Correlation functions are off-shell objects and their one-particle irreducible (1PI) parts 
in a given background are given by derivatives of the effective action with respect to the fields. While the respective right hand side of 
\labelcref{eq:FunDSE} does not vanish, the evaluation on the EoM minimises the loop contributions.  These one-particle irreducible (1PI) parts of the correlation functions of QCD are given by 
\begin{align}
\langle \hat \Phi_{i_1}(p_1)\cdots \hat\Phi_{i_n}(p_n)\rangle^{\ }_\textrm{1PI}=	\Gamma^{(n)}[\Phi_\textrm{EoM} ](p_1,...,p_n)\,, 
\end{align}
where $\Gamma^{(n)}$ stands for the $n$th derivative of the effective action with respect to the fields $\Phi_{i_1},...,\Phi_{i_n}$. In the vacuum, the global minimum of the effective action is given by $(\bar A, \Phi)_\textrm{EoM}=0$. This trivial background is the commonly used background for the computation of correlation functions in functional approaches, and in particular for the DSE and fRG approaches. 

At finite temperatures, the confining dynamics manifests itself in the non-trivial expectation value of the temporal component of the gauge field, tantamount to including the dynamics of the Polyakov loop. Its effective potential is readily computed in functional QCD, \cite{Braun:2007bx, Fister:2013bh}, and applications to the phase structure at real and imaginary potential can be found in \cite{ Braun:2009gm, Fischer:2013eca, Fischer:2014vxa, Fischer:2014ata}, see also \cite{Reinosa:2014ooa, Reinosa:2015oua, Maelger:2017amh, 10.21468/SciPostPhys.12.3.087, vanEgmond:2024ljf, MariSurkau:2025pfl} for applications in the Curci-Ferrari model. This background, that signals the confining dynamics, is key to the description of the transition from the quark-gluon phase to the hadronic one. In the present work we store this information completely in the background $\bar A_0\neq 0$, while keeping $a_0=0$, for other options see e.g.~\cite{10.21468/SciPostPhys.12.3.087, vanEgmond:2024ljf}. 

Importantly, in a trivial background with $\bar A_0=\bar a_0=0$, the low temperature cumulants do not even describe the qualitative features of the hadronic phase. This regime includes the transition regime around $T_c$. For a comprehensive analysis see \cite{Fu:2015gl}.  There, it has been shown in particular, that the kurtosis does not even get close to unity for vanishing temperature in a trivial background, and a unity value of the kurtosis signals the hadronic phase with weakly interacting hadrons.  More recently, it has also been shown that the confining background is crucial for a quantitative description of the isentropic trajectories and other thermodynamic observables~\cite{Lu:2023msn}.  Related works in the fRG within quantitative QCD-assisted low energy effective field theories are provided in \cite{Fu:2016tey, Fu:2021oaw, Fu:2023lcm}, and the latter two are based on the functional QCD work on the phase structure of QCD in \cite{Fu:2019hdw}. The present DSE work in the fQCD collaboration \cite{fQCD} will be paired with an fRG one in preparation.

%%%%%%%%%%%%%%%%%%%%%%%%%%%%%
\subsection{QCD correlation functions}
\label{sec:QCDCorrelations} 

In a first step we compute correlation functions in 2+1 flavour QCD from their coupled set of DSEs in a vanishing background, specifically the quark and gluon propagators. The computational details can be found in \cite{Gao:2020qsj, Gao:2020fbl, Gao:2021wun}, and we only briefly summarise the important ingredients: we use the miniDSE scheme put forward in \cite{Lu:2023mkn}: in this setup thermal and chemical potential corrections of QCD correlation functions are computed from their difference DSEs. The reliability of this scheme crucially depends on the quantitative precision of the vacuum QCD input and we 
use the 2+1 flavour QCD data for quark and gluon propagator as well as for all tensor structures of the quark-gluon vertices obtained from the precision DSE computation \cite{Gao:2021wun} in the vacuum. For the ghost propagator we use the two flavour data from the precision fRG computation in \cite{Cyrol:2017ewj}. This direct use of the fRG results is possible as we use the fRG-compatible renormalisation MOM${}^2$ scheme set up in \cite{Gao:2021wun}. We neglect the dependence of the ghost propagator on the strange quark as well as on $T,\mu_B$ which has been proven top be subleading. Moreover, the miniDSE systematics allows us to use the STI- and RG-adapted quark-gluon vertex dressings derived in \cite{Gao:2020qsj}. The quantitative accuracy of these dressings in the phase structure of QCD has been checked in \cite{Gao:2020fbl}: the location of the chiral crossover lines agree which each other, while the $\mu_B$-location of the critical end point in the present approximation is 10\% larger than in \cite{Gao:2020fbl}. In the latter work it has been argued, that the full quantitative reliability of the current approximation and respective functional renormalisation group works, for a discussion see \cite{Fu:2019hdw, Fu:2023lcm}, is limited by $\mu_B/T\lesssim 4$. In turn, for $\mu_B/T\gtrsim 4$, the functional results represent estimates with successively larger systematic errors. However, the clustering of the functional results for the CEP $600 \lesssim \mu_B\lesssim 650$\,MeV for different approaches and resummations reduces this systematic error considerably. In short, the current approximation captures the full QCD dynamics computed in the present state of the art quantitative DSE studies with a significantly smaller computational effort for $\mu_B/T\lesssim 4$, and the results for $\mu_B/T\gtrsim 4$ should provide good QCD estimates.

%%%%%%%%%%%%%%%%%%%%
\section{Phase structure of QCD}
\label{sec:PhaseStructure}

In this Section we map out the phase structure of QCD and access both, the confinement-deconfinement and chiral crossovers. The confinement-deconfinement crossover and the respective order parameter is discussed in \Cref{sec:Conf-Deconf}. Specifically, we compute the non-trivial temperature and baryon chemical potential dependence of the non-trivial gauge invariant gluonic background, that signals the confining dynamics in the hadronic phase. Apart from allowing the access to the   confinement-deconfinement crossover, it also is of crucial importance for the results for thermodynamic observables and the fluctuations of conserved charges in the hadronic and crossover regimes. In \Cref{sec:ChiralPhaseTransition} we compute the chiral condensate and the chiral crossover temperature from its thermal susceptibility. We also show that the chiral condensate is essentially independent of the gluonic background. Finally, in \Cref{sec:PhaseStructureResults} we put everything together and present our results for the phase structure.

%%%%%%%%%%%%%%%%%%%%
\subsection{Confinement-deconfinement phase transition}
\label{sec:Conf-Deconf}

In this Section, we access the confinement-deconfinement crossover with an order parameter observable for center-symmetry breaking. 
This order parameter is provided by the gauge invariant eigenvalue field $\nu$ of the algebra element of the Polyakov loop, see \Cref{app:PloopAlg}. 
Obviously, it is related to the standard order parameter for center-symmetry breaking, the expectation value of the traced Polyakov loop, but is more amiable towards its computation in functional approaches. 
Moreover, it does not suffer from the presence of normalisation factors with a large temperature dependence which diffuses the access to the crossover temperatures in the case of the expectation value of the traced Polyakov loop, see \cite{Herbst:2015ona}. 
This order parameter has been suggested and worked out within the fRG approach in \cite{Braun:2007bx, Marhauser:2008fz, Braun:2010cy} for different gauge groups and gauges. 
Its formulation within general functional approaches has been put forward in \cite{Fister:2013bh}, and its relation to the commonly used order parameter, the expectation value of the traced Polyakov loop, has been evaluated in \cite{Herbst:2015ona}. For a formulation with manifest center symmetry see also~\cite{10.21468/SciPostPhys.12.3.087, vanEgmond:2024ljf}. 

The details of the functional approach to the Polyakov loop or rather the eigenvalue field $\nu$ are provided in  \Cref{app:Poloop+PolPot}. The understanding of the approach is chiefly important for that of the computations in the present work, and hence for an evaluation of the underlying systematics. For that purpose and for the benefit of the technically interested reader we provide a brief review of this approach in \Cref{app:Poloop+PolPot}: In \Cref{app:PloopAlg} we review the construction of the gauge invariant functional order parameter and its (linear) relation to the expectation value of the temporal gauge field in the present setup. We also highlight its importance as an optimal background or expansion point for the systematic expansion schemes in functional approaches. In \Cref{app:DSE-Polya} we provide a step-by-step analysis of the computational details. 

In the present Section we utilise the functional Polyakov loop approach detailed in \Cref{app:Poloop+PolPot} for an evaluation of the 
confinement-deconfinement crossover, or rather that of center-symmetry breaking, in the phase structure of QCD.  
We provide the order parameter potential $V_\varphi(\varphi_3,\varphi_8)$, where $\varphi_{3,8}$ are related to the Cartan component of the temporal gauge field, 
$\varphi_{3,8} = 2 \pi A_0^{3,8} /( g_s \beta )$, see \labelcref{eq:varphiA0}. As discussed in \Cref{app:DSE-Polya}, the order parameter potential $V_\varphi$ is computed from a sum over eigenvalues and modes of the mode potentials \labelcref{eq:VmodeComp}. 
In the present approximation, the only non-trivial input in the latter are the spatial-momentum and frequency dependent dressings $Z_{E,M}, Z_c, Z^{E,M}_q, M_q$ of the full ghost, gluon and quark propagators, see \labelcref{eq:ModeIntegrandsfa,eq:ModeIntegrandsfc,eq:ModeIntegrandsfq}. The computational details have been explained in the introduction of \Cref{sec:QCDCorrelations}. 
Here we only add the details concerning the quark-gluon vertex. 
As mentioned in the introduction, we resort to the computational scheme in \cite{Gao:2020qsj} for the dominant vertex dressings $\lambda_{q\bar{q}A}^{1,4,7}$. 
In  \cite{Gao:2020qsj}, a combination of the  Slavnov-Taylor identities (STIs) as well as RG-invariance has been used for the construction of STI+RG--consistent dressings. 
The quantitative reliability of this setup has been confirmed in \cite{Gao:2020fbl}, where the dressings of the quark-gluon vertex have been computed directly. In conclusion the results from the present scheme agree with that of the full computation in \cite{Gao:2020fbl} up to chemical potentials $\mu_B\approx 600\,$MeV. Note that this is already beyond the full quantitative reliability of the computation in  \cite{Gao:2020fbl} of $\mu_B/T\lesssim 4$, see also \cite{Fu:2019hdw} and the comprehensive analysis in \cite{Fu:2023lcm}. 
Moreover, we stabilise the setup in \cite{Gao:2020qsj} further by using the  vacuum 2+1-flavour QCD input from~\cite{Gao:2021wun} instead of the two-flavour input from  \cite{Cyrol:2017ewj}. 
This entails that we do not have to accommodate the strange quark dependence in the difference DSE. 
A further crucial benchmark is the compatibility of the results for the chiral crossover temperature $T_c(0)$ at vanishing chemical potential and the curvature $\kappa_2$ of the crossover line at $\mu_B=0$. 
We find 
\begin{align} 
	T_c(0)=157\,\textrm{MeV}\,,\qquad \kappa_2 = 0.0153\,, 
\label{eq:Tchiral0}
\end{align}
which is in quantitative agreement with the lattice results as well as that from quantitative functional computations, see \cite{Fu:2019hdw, Gao:2020qsj, Gao:2020fbl, Gunkel:2021oya}. 
Further, qualitative, functional studies typically fail to meet the curvature constraint, for recent compilations of the respective results see \cite{Gao:2020qsj, Bernhardt:2023ezo}. 
In summary, the sophisticated approximation used here for the coupled DSEs of the quark and gluon propagators dressings as well as that of the quark-gluon meets all the benchmarks at $\mu_B=0$. The location of the critical end point is obtained at 
\begin{align}
(T^{\textrm{CEP}},\mu_B^{\textrm{CEP}})=(103,660)\,\textrm{MeV}\,. 
\label{eq:LocationCEP}
\end{align}
\Cref{eq:LocationCEP} agrees within the errors with the up-to-date estimates ~\cite{Fu:2019hdw,Gao:2020fbl,Gunkel:2021oya}. 
We note in passing that \labelcref{eq:LocationCEP} should not be understood as a new estimate. 
The state-of-the-art estimate is still given by \cite{Gao:2020fbl}, which is singled out by its direct computation of the quark-gluon vertex dressing. 
We hasten to add that an update of the estimate is not one of the aims of the present work. 
Here we aim at the computation of thermodynamic observables including the first quantitative DSE computation of the fluctuations of conserved charges. 

\begin{figure}[t]
	\centering
	\includegraphics[width=0.95\columnwidth]{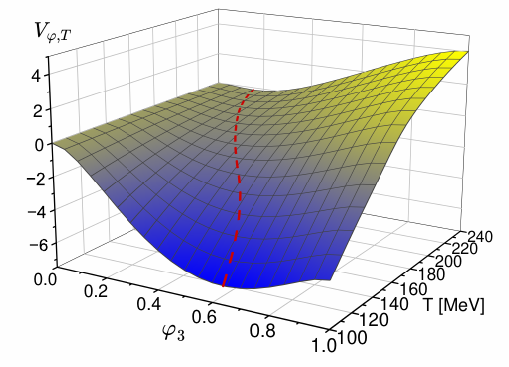}
	\caption{3D plot of the dimensionless Polyakov loop potential $V_{\varphi,T}(\varphi_3,0)$, defined in \labelcref{eq:DeltaV,eq:DefofVvarphi}. We show it as a function of the background field $\varphi_3$ and the temperature $T$ at $\varphi_8=0$ and vanishing baryon chemical potential. The temperature-dependent minimum of the Polyakov loop potential is displayed as a dashed red line.}
	\label{fig:VphiT0}
\end{figure}
The above analysis of the results for the chiral phase structure concludes the discussion of present setup: 
we have quantitative reliability up to $\mu_B/T\lesssim 4$ and agree with the state-of-the-art DSE computation in \cite{Gao:2020fbl} up to baryon chemical potentials $\mu_B\approx 600$\,MeV, very close to the critical end point.  

We proceed with the results for the confinement-deconfinement or rather center phase transition in QCD. 
We have discussed in \Cref{app:Poloop+PolPot}, that the expectation values of the gauge invariant eigenvalues $\langle \nu_\pm\rangle$ and $\langle \nu_3\rangle$ determine the expectation values $\langle\varphi_3\rangle, \langle\varphi_8\rangle$, \labelcref{eq:Lnu2}. 
Hence, the latter are gauge invariant observables and constitute order parameters for the confinement-deconfinement phase transition, or rather the center phase transition in QCD. 
The expectation values $\langle \varphi_{3,8}\rangle $ are simply the solutions of the equations of motion of $\varphi_{3,8}$ which are the saddle points of the Polyakov loop potential, see \labelcref{eq:EoMVvarphi}. Accordingly, we can read off their values from these saddle points. 

\begin{figure}[t]
	\centering
	\includegraphics[width=0.92\columnwidth]{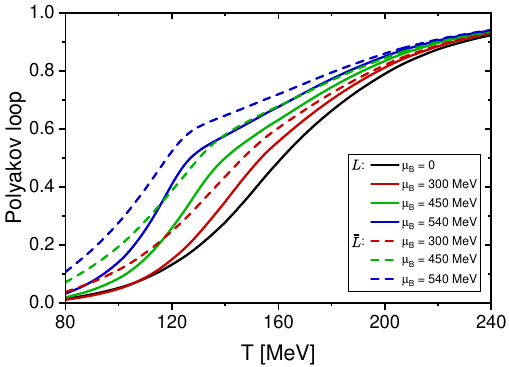} 
	\caption{Traced Polyakov loops ${L}$ and $\bar{L}$ of the expectation value of the eigenvalue field, $\langle \nu\rangle $, see \labelcref{eq:barvarphi38Main,eq:LbarL-muqMain}. They are shown as a function of temperature $T$ for different values of baryon chemical potential $\mu_B$. The respective thermal susceptibilities \labelcref{eq:LbarLSusceptibilities} are shown in \Cref{fig:suscL-app}.}
	\label{fig:LoopTmu}
\end{figure}
At vanishing chemical potential the physical solution of the EoM is given by the absolute minimum of the potential. Moreover, for $\mu_B=0$, the expectation value of $\varphi_8$ is vanishing for all $T$, a respective heat map of the potential for  $T=120$\,MeV is shown in \Cref{fig:V38_contour_120_0} in \Cref{app:DSE-Polya} for illustration. This entails that we only have to consider the temperature dependence of the potential in the $\varphi_3$ direction. This is depicted in \Cref{fig:VphiT0}, and we have normalised the potential such that it vanishes at vanishing background field, 
\begin{align}
  V_{\varphi,T}(\varphi_3,\varphi_8) = \left. V_\varphi(\varphi_3,\varphi_8)\right|_T - \left. V_\varphi(0,0)\right|_{T},
\label{eq:DeltaV}
\end{align}
The minimum $\langle \varphi_3\rangle$ is indicated by a dashed red line in \Cref{fig:VphiT0}. 
For asymptotically large temperatures, $T\to \infty$, it vanishes, $\langle \varphi_3\rangle\to 0$. In turn, for $T\to 0$ it approaches the $\langle \varphi_3\rangle=2/3$ with $L(\langle \varphi_3\rangle , 0)=0$. 
The expectation values $\langle \varphi_{3,8}\rangle $ are the gauge invariant order parameters that serve as optimal backgrounds for most expansion schemes in functional approaches and specifically for the vertex expansion that underlies most computations. However, the confinement-deconfinement or center symmetry properties are more conveniently assessed  with the Polyakov loop of the expectation value of the eigenvalue field, $L(\langle \nu\rangle )$  and $\bar L(\langle \nu\rangle )$, see \labelcref{eq:LbarL-muq} in \Cref{app:DSEA0}. For convenience and their importance we recall the relations here 
\begin{align}\nonumber 
	L(\langle \nu\rangle ) =&\, \frac{1}{3}\left[ e^{ 2 \pi \bar{\varphi}_8/\sqrt{3}} + 2 \, e^{-\pi \bar{\varphi}_8/\sqrt{3}} \cos{\pi\bar \varphi_3}\right]\,, \\[1ex] 
	\bar L(\langle \nu\rangle )  = &\,\frac{1}{3} \left[e^{- 2 \pi \bar{\varphi}_8/\sqrt{3}} + 2 \, e^{\pi \bar{\varphi}_8/\sqrt{3}} \cos{\pi\bar \varphi_3}\right]\,,  
	\label{eq:LbarL-muqMain}
\end{align}
with 
\begin{align}
\varphi_3=\langle \hat \varphi_3\rangle \,,\qquad  \bar \varphi_8= -\imag  \langle \hat \varphi_8\rangle \,,
\label{eq:barvarphi38Main}	
\end{align}
which are related to the expectation value of the eigenvalue field, 
\begin{align}
	\langle \nu_{\pm}\rangle = \frac12 \left( \imag \bar \varphi_8 \pm \frac{1}{\sqrt{3}}\, \bar \varphi_3\right)\,,\qquad \langle \nu_3\rangle = -\imag \bar \varphi_8 \,, 
	\label{eq:nuFieldExpMain}
\end{align}	
The observables \labelcref{eq:LbarL-muqMain} are depicted in \Cref{fig:LoopTmu} as a function of the temperature for different chemical potentials $\mu_B=0$. We define the confinement-deconfinement crossover temperatures $T_L(\mu_B)\,,\,T_{\bar L}(\mu_B)$ with the peaks of the thermal susceptibilities of the Polyakov loops $L,\bar L$ in \labelcref{eq:LbarL-muqMain}, 
\begin{align}
  \chi^{\ }_{{L}} = \frac{\partial{{L}}}{\partial T}\,,\qquad  \chi^{\ }_{{\bar L}} = \frac{\partial{{\bar L}}}{\partial T}\,. 
\label{eq:LbarLSusceptibilities}
\end{align}
\begin{figure}[t]
	\centering
	\includegraphics[width=0.9\columnwidth]{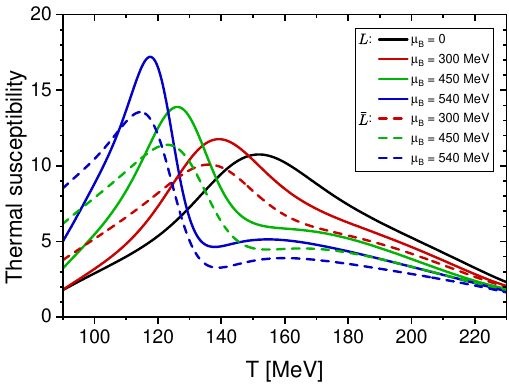} 
	\caption{Thermal susceptibilities \labelcref{eq:LbarLSusceptibilities} of the Polyakov loops $L,\bar L$, see \labelcref{eq:barvarphi38Main,eq:LbarL-muqMain}. They are shown as functions of temperature $T$ for different values of baryon chemical potential $\mu_B$, matching that in \Cref{fig:LoopTmu}. }
	\label{fig:suscL-app}
\end{figure}
The susceptibilities \labelcref{eq:LbarLSusceptibilities} are shown in \Cref{fig:suscL-app} for the same chemical potentials as used  in \Cref{fig:LoopTmu}. We define the thermal widths of the susceptibilities by the boundary of the interval in which $L,\bar L$ exceed half their peak height. This coincides roughly with the width of a Breit-Wigner fit about the peak. In particular, we find at vanishing chemical potential
\begin{align}
	T_{{L}}(0) = T_{\bar {L}}(0) = 152\,\textrm{MeV}\,,
\end{align}
close to the pseudo-critical temperature of the chiral transition $T_c(0) = 157\,\textrm{MeV}$, see \labelcref{eq:Tchiral0}.  

Our results for the confinement-deconfinement crossover are shown in \Cref{fig:PD-conf-width} in \Cref{sec:ChiralPhaseTransition} together with the chiral crossover temperature and its width.

%%%%%%%%%%%%%%%%%%%%%%%%%%
\subsection{Chiral phase transition}
\label{sec:ChiralPhaseTransition}

We proceed with the evaluation of the chiral order parameter, the renormalised chiral condensate $\Delta_{l,R}$ in \labelcref{eq:cond-reduced}, and the respective crossover temperature $T_\chi$. Here we follow \cite{Fu:2019hdw,Gao:2020qsj,Gao:2020fbl}, and the only novel aspect is the discussion of the dependence of $\Delta_{l,R}$ on the gluonic background $\varphi$, or rather the lack thereof. 

The renormalised light chiral condensate $\Delta_{l,R}$ reads 
\begin{align}
	\Delta_{l,R} = \frac{2}{m_{\pi}^4} \sum_{q=u,d} \left[ \Delta_q(T,\mu_B) - \Delta_q(0,0) \right]\,.
	\label{eq:cond-reduced}
\end{align}
\Cref{eq:cond-reduced} is the difference between the chiral condensate $\Delta_q$ at $(T,\mu_B)\neq 0$ and that in the vacuum, with   
\begin{align}
	\Delta_q = -m_{q} T \sum_{\omega_p} \int \frac{\mathrm{d}^{3}\spatial{p}}{(2\pi)^{3}} \, \tr\, [ G_{q\bar q}(p_q) ] \,, 
	\label{eq:qbarq-A0}
\end{align}
where $m_q$ is the current quark mass.  
The trace $\tr$ sums over colour and Dirac indices and the momentum and frequency variable $p_q$ is given by 
\begin{align} 
	(p_{q,\mu}) = (\omega_{q,n} + 2 \pi \varphi^c- \imag \mu_q \,,\,\boldsymbol{p})\,, 
	\label{eq:pqMain}
\end{align}
with the quark chemical potential $\mu_b=\mu_B/3$ and the background field $\varphi^c$ in the Cartan subalgebra with the components $\varphi_{3,8}$, see \labelcref{eq:varphiA0}. 
\Cref{eq:qbarq-A0} simply is the integrated scalar part of the quark propagator. The full quark propagator is parametrised as 
\begin{align}
	G_{q\bar q}(p)  
	=&\, \Bigl[-\imag  \gamma_0  p_{q,0}-\frac{Z_q^M}{Z_q^E} \imag  \boldsymbol{\gamma}\, \boldsymbol{p}  +M_q \Bigr] \, f_q(p_q,\mu_q) \,,
	\label{eq:Gqbarq}
\end{align}
with the scalar dressing $ f_q$,  
\begin{align}
	f_q(p_q,\mu_q)=  \frac{1}{Z_q^E} \frac{1}{  p_{q,0}^2+\left(\frac{Z_q^M}{Z_q^E}\right)^2\boldsymbol{p}^2+M^2_q} \,, 
	\label{eq:ModeIntegrandsfqMain}
\end{align}
and the momentum, chemical potential and background-dependent dressings $Z_q^{E,M}(p_q,\mu_q)\,,\, M_q(p_q,\mu_q)$. These dressings are computed from the quark gap equation. They are complex functions of $p_q$ and $\mu_q$ and we have deferred the discussion of the computational details to \Cref{app:Vq}. Moreover, as discussed there in the context of the computation of the  the quark part of the Polyakov loop potential, the colour part of the trace in \labelcref{eq:qbarq-A0} can be turned into a sum over the eigenvalues of $\varphi^c$. 

\begin{figure}
	\centering
	\includegraphics[width=0.93\columnwidth]{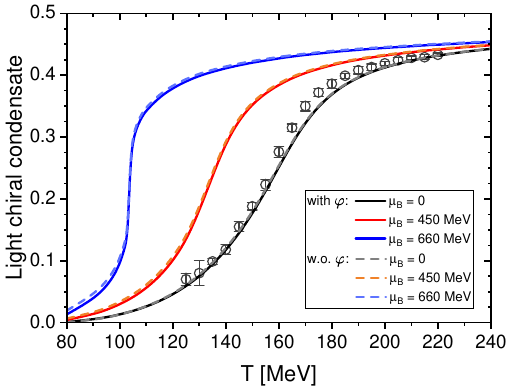}
	\caption{Renormalised light chiral condensate $\Delta_{l,R}$, \labelcref{eq:cond-reduced}, as a function of the temperature $T$ for several chemical potentials $\mu_B$ including $\mu_B^{\textrm{CEP}}=660\,$MeV at the critical end point: $\Delta_{l,R}$ with the gluonic background $\varphi$ (straight), $\Delta_{l,R}$ at $\varphi=0$ (dashed). We also show the lattice QCD result of $\Delta_{l,R}$ at zero chemical potential (open circles) from \cite{Borsanyi:2010bp}. For $\Delta_{l,R}$ with the gluonic background $\varphi$, the respective thermal susceptibilities \labelcref{eq:DeltaTRSusceptibilities} are shown in \Cref{fig:suscChi}.} 
	\label{fig:condA0}
\end{figure}
The chiral crossover temperature is computed from the thermal susceptibility, 
\begin{align} 
\chi_{\Delta_{T,R}} = \frac{\partial 	\Delta_{T,R} }{\partial T}\,, 
	\label{eq:DeltaTRSusceptibilities}
\end{align}
see e.g.~\cite{Gao:2020fbl, Gao:2020qsj, Gunkel:2021oya} for previous computation with DSEs. 
For a comparison of crossover temperatures from different order parameters see \cite{Pawlowski:2014zaa,Fu:2019hdw}. 
The respective crossover temperature is shown in \Cref{fig:PD-conf-width} together with its width. 
The latter is  defined by the boundaries of the interval in which the susceptibility exceeds half its peak height, matching the definition of the confinement-deconfinement crossover. 

\begin{figure}[b]
	\centering
	\includegraphics[width=0.9\columnwidth]{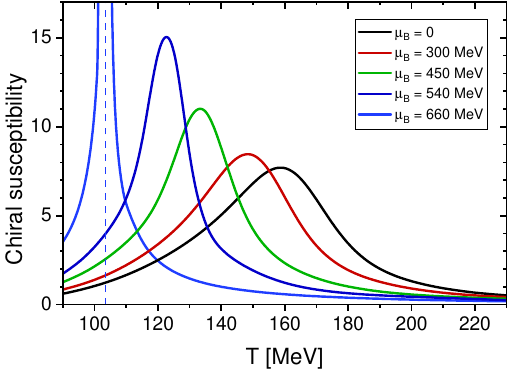}
	\caption{ Thermal susceptibility of the light chiral condensate $\chi_{\Delta_{l,R}}$, \labelcref{eq:DeltaTRSusceptibilities}, as a function of the temperature $T$ for several chemical potentials $\mu_B$ including $\mu_B^{\textrm{CEP}}=660\,$MeV at the critical end point. The susceptibility is shown as a function of temperature $T$ for different values of baryon chemical potential $\mu_B$, matching that in \Cref{fig:condA0}. }
	\label{fig:suscChi}
\end{figure}

In summary, the chiral condensate $\Delta_q$ is computed in the  gluonic background $\varphi$ and are shown in \Cref{fig:condA0}. 
Interestingly, the difference between the results in the gluonic background and that with a vanishing background ($\bar A_0=0$) are negligible. 
The non-trivial background is triggered by the confining dynamics and reflects the latter phenomenon. 
This suggests that the onset of confinement has little impact on the size of dynamical chiral symmetry breaking. 
This is specifically interesting, as it is qualitatively different for the fluctuation of conserved charges. 
The latter depend qualitatively on the `confining' background, for a comprehensive analysis see \cite{Fu:2015naa, Fu:2016tey}. We shall corroborate the findings there in \Cref{sec:ResultsObservables}, where thermodynamic observables and the fluctuations of conserved charges are computed within functional QCD. 

We close this Section with the remark, that have used the quotation marks as the background is not confining but rather signals confinement. We will still keep the commonly used notion of `confining' and its true meaning is implicitly understood.

%%%%%%%%%%%%%%%%%%%%
\subsection{QCD  Phase structure}
\label{sec:PhaseStructureResults}

\begin{figure}[t]
\centering
\includegraphics[width=0.9\columnwidth]{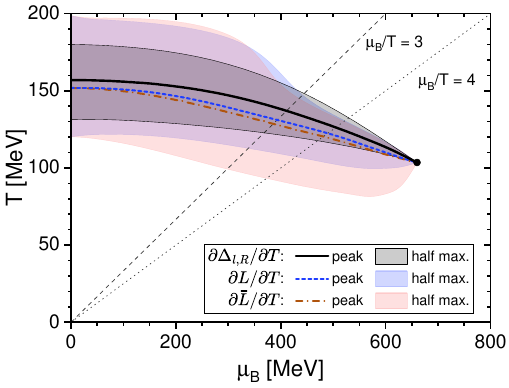} 
\caption{Chiral and confinement-deconfinement crossover temperatures $T_\chi,T_L,T_{\bar L}$ as functions of the baryon chemical potential. They are defined by the peaks of the thermal susceptibilities: $\partial \Delta_{l,R}/\partial T$ (straight black, width: grey band), $\partial {L} / \partial T$ (dashed blue, width: blue band) and $\partial {\bar L}^{} / \partial T$ (dash-dotted red, width: red band). The width of the crossovers are defined by the width at half maximum of the thermal susceptibilities. }
\label{fig:PD-conf-width}
\end{figure}

The results of \Cref{sec:Conf-Deconf,sec:ChiralPhaseTransition} allow us to map out the chiral and confinement-deconfinement phase structure of QCD at vanishing strangeness and charge, \labelcref{eq:muqs}. 
In \Cref{fig:PD-conf-width} we show the chiral and confinement-deconfinement crossover lines together with their thermal widths.  
As expected, the chiral crossover is far steeper than the confinement-deconfinement ones, and the $\bar L$-crossover is the smoothest one. 
Still, in contradistinction to the expectation value of the Polyakov loop as measured with lattice simulation, the confinement-deconfinement crossover temperatures can be determined well and they are close to the chiral one. This originates in the fact that they are not subject to large thermal normalisation factors, see \cite{Herbst:2015ona}. The latter shift the transition temperature to far larger values. 

The physics significance of the respective confinement-deconfinement temperatures is supported by the following property: the gauge invariant order parameters $\langle \varphi_{3,8}\rangle$ play an important physics rôle as physical backgrounds that allow for optimal convergence of computations of observables within functional approaches. Loosely speaking, such an expansion point already carries a sizeable part of the underlying dynamics. 
Indeed, resorting to the diagrammatic representation of the Polyakov loop potential, the expectation values themselves have a diagrammatic representation. 
If inserting them as backgrounds in the diagrammatic representation of, e.g., the fluctuations of conserved charges, this background provides an additional, physically relevant, resummation. In summary this leads to a qualitatively enhanced apparent convergence of such a computation. 

This leads us directly to the assessment of the systematic error of the phase structure predictions within the current approximation. 
This is important for the assessment of the systematic error of the results for the thermodynamics and the fluctuations of conserved charges in \Cref{sec:ResultsObservables}: To begin with, the present results for the chiral phase transition lines, including the location of the CEP, are in quantitative agreement with the functional results in \cite{Fu:2019hdw, Gao:2020fbl, Gunkel:2021oya} and the closely related one \cite{Gao:2020qsj}. Indeed, the present computation is an upgrade of the latter computation and this upgrade is informed by the improvements on \cite{Gao:2020qsj} in of the state-of-the-art DSE computation in \cite{Gao:2020fbl}. This entails that its respective systematic error can be deduced from \cite{Gao:2020fbl}. 

\begin{figure}[t]
	\centering
	\includegraphics[width=0.9\columnwidth]{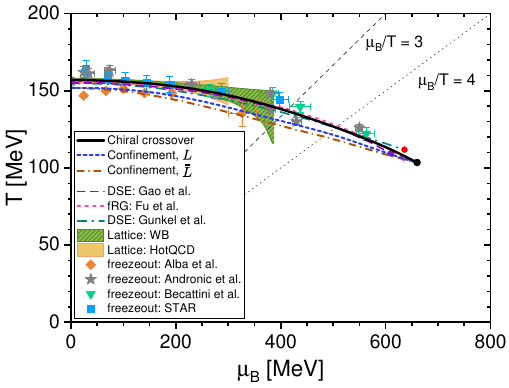}
	\caption{Chiral and confinement-deconfinement phase structure of QCD. The crossover lines display the locations of the peak positions of the thermal susceptibilities $\chi_{\Delta_{T,R}}$ (chiral,\labelcref{eq:DeltaTRSusceptibilities})  $\chi_L$, $\chi_{\bar L}$ (confinment-deconfinement, \labelcref{eq:LbarLSusceptibilities}). We also show the chiral phase transition line obtained from the lattice QCD~\cite{Borsanyi:2020fev, HotQCD:2018pds} and functional QCD approaches~ \cite{Fu:2019hdw, Gao:2020fbl, Gunkel:2021oya}, and the freeze-out data from different groups~\cite{Alba:2014eba, Andronic:2017pug, Becattini:2016xct, STAR:2017sal}. 
	}
	\label{fig:PhaseStructureChiralConf2025}
\end{figure}

For the discussion of the combined systematic error of the theoretical computations of the QCD phase structure we use all state-of-the-art results both from lattice simulations as well as functional approaches of QCD.s In \Cref{fig:PhaseStructureChiralConf2025} we show the respective chiral and confinement-deconfinement crossover lines as well as freeze-out data from different groups. In both Figures, \Cref{fig:PD-conf-width,fig:PhaseStructureChiralConf2025}, we indicate the two lines $\mu_B/T = 3$ and $\mu_B/T=4$ relevant for the systematic error analysis: The former one, $\mu_B/T = 3$, indicates the convergence area of lattice simulations. Beyond this regime, with  $\mu_B/T \gtrsim 3$, one may use extrapolations of lattice results. However, the respective unbiased systematic error estimate is increasing rapidly and only can be kept small by additional assumptions such as the absence of novel physics phenomena such as (off-shell) diquark and baryon fluctuations or the dynamics of the density mode.  In conclusion, lattice simulations at $\mu_B=0$ offer benchmark results for functional computations at vanishing chemical potentials and  functional computations as first principle QCD ones, that, amongst other criteria,  meet all these benchmark results. For a systematic discussion of these properties in the phase structure of QCD we refer the reader to \cite{Fu:2023lcm}, for general discussions of apparent convergence in functional approaches see \cite{Ihssen:2024miv, Lu:2023mkn, Fu:2025hcm, Huber:2025kwy}

Consequently, for $\mu_B/T \gtrsim 3$, the only QCD benchmark results to date are the functional ones from \cite{Fu:2019hdw, Gao:2020fbl, Gunkel:2021oya}. Then, the systematic error line at $\mu_B/T=4$ indicates the current regime of quantitative reliability of functional approaches, both from state-of-the art fRG and DSE computations \cite{Fu:2019hdw, Gao:2020fbl, Gunkel:2021oya}, see \Cref{fig:PhaseStructureChiralConf2025}. 
In contradistinction to the lattice line at $\mu_B/T=3$ this does not come as a hard conceptual bound.
 Indeed, we hope to report on more sophisticated computations soon, where this bound is pushed beyond the regime of the critical end point or rather that of the onset of new phases: the direct systematic error estimate of the different functional computations in \cite{Fu:2019hdw, Gao:2020fbl, Gunkel:2021oya} increases successively for $\mu_B/T\gtrsim 4$. 
 This increase of the systematic error estimate is specifically triggered due to the onset of the moat regime \cite{Pisarski:2021qof} in QCD, \cite{Fu:2019hdw, Fu:2024rto} and its incomplete resolution in the current state-of-the-art functional computations. 
 This deficiency is currently resolved and \cite{Fu:2024rto} is a first important step in this direction. 
 Still, we remark that the quantitative agreement of the different functional approaches in the regime $\mu_B/T\gtrsim 4$ provides non-trivial evidence for the reliability of the results even in this regime. 
 This remarkable agreement even persists for the location of the CEP which only varies within 10\%. 

This concludes our discussion of the systematic error estimate. In short, the results for the thermodynamics and fluctuations of conserved charges have a small systematic error for $\mu_B/T\lesssim 4$. For $\mu_B/T\gtrsim  4$, the direct systematic error is increasing successively, but the agreement of different functional computation within rather different resummations schemes suggest a still rather small systematic error of 10\%.  

Finally we remark that our present framework successfully captures the relative $\mu_B$- shifts of $L$ and $\bar{L}$, see \Cref{fig:LoopTmu}. Their difference is related to the free energy gap between a static quark and an anti-quark. This further allows for a systematic study on strangeness neutrality~\cite{Fukushima:2009dx,Rennecke:2019dxt} via continuum correlation functions, which will be presented in a follow-up work~\cite{FlucsfRG-QCD}.

%%%%%%%%%%%%%%%%%%%%%%%%%%%%%%%%%%%%%%
\section{Thermodynamics and fluctuations of conserved charges}
\label{sec:ResultsObservables}

The preparations and results of \Cref{sec:PhaseStructure} and \Cref{app:Poloop+PolPot} for the chiral and confinement-deconfinement crossovers allow us to resolve the thermodynamics and fluctuations of conserved charges with DSEs both quantitatively. 
As already alluded to \Cref{sec:ChiralPhaseTransition}, the gluonic background $\varphi$ (or $A_0$), triggered by the confining dynamics, is essential for even qualitative results. In particular, they are of crucial importance for the fluctuations of the conserved net-baryon number, see \cite{Fu:2015naa,Fu:2016tey}. 
Moreover, this importance is even growing for higher order fluctuations. 
The current results are benchmarked with the lattice results at vanishing chemical potential. 
For $\mu_B/T\gtrsim 3$ they are discussed in comparison to the fRG results in \cite{Fu:2021oaw, Fu:2023lcm} within sophisticated low energy effective theories: they pass the benchmark tests at $\mu_B=0$ and corroborate the fRG results at  $\mu_B\neq 0$. We also emphasise that, both the present results and \cite{Fu:2021oaw, Fu:2023lcm} confirm the findings in \cite{Fu:2015naa, Fu:2016tey}. While the chiral properties, such as the renormalised chiral condensate $\Delta_{T,R}$, show a relatively small dependence on the confining background $\langle \varphi\rangle $ , the fluctuations of conserved charges in QCD show a qualitative dependence. In short, the results in a vanishing background are not converged and fail to even reflect the qualitative dependence in QCD already at $\mu_B=0$. In turn, in the confining background they agree quantitatively with the lattice results at $\mu_B=0$ and with each other for larger densities, and in particular for $\mu_B/T \gtrsim 3$. 

In \Cref{sec:baryon-flucs}, we discuss our results for the baryon number density and the fluctuations of conserved charges. 
The respective results for the baryon number kurtosis are used in \Cref{sec:eos-A0-bes} for a discussion of the beam energy scan (BES).

%%%%%%%%%%%%%%%%%%%%%%%%%%%%%%%%%%%
\subsection{Baryon number density and thermodynamics}
\label{sec:nB+Thermodynamics}

The baryon number density and its respective fluctuations are readily derived from the net-quark number density, 
\begin{align}
n_q(T,\mu_q) \simeq  T \sum_{\omega_p} \int \frac{\mathrm{d}^{3}\spatial{p}}{(2\pi)^{3}} 
\langle\bar{q}(-p)\gamma_0\,q(p)\rangle\,,
\label{eq:nqDef}
\end{align}
where we have left out a (re-)normalisation function, for a respective discussion see e.g.~\cite{Lu:2023mkn, Gao:2021nwz}. The net-quark number density is given by the trace of the $\gamma_0$-component of the quark propagator,  
\begin{align}
n_{q}(T,\mu_q) = - T \sum_{\omega_p} \int \frac{\mathrm{d}^{3}\spatial{p}}{(2\pi)^{3}} \tr\,\bigl[ \gamma_0\, \bar G_{q}(p_q)  \bigr]\,, 
\label{eq:nq}
\end{align}
with $p_q=p_q(\varphi^c)$ defined in \labelcref{eq:pqMain}. The gluonic background $\varphi^c$ takes care of the fact that the quark propagator in \labelcref{eq:nqDef} is the one on the full equations of motion. The trace $\tr$ in \labelcref{eq:nq} sums over colour and Dirac indices, and the RG-invariant propagator $\bar G_q$ is defined by  
\begin{align} 
	 \bar G_{q}(p_q) = Z_q^{E}(p_q)\, G_{q\bar q}(p_q) \,,
	 \label{eq:barGq}
	\end{align}
The RG-invariance of $\bar G_{q}$ is accommodated by the wave function $Z_q^{E}(p_q)$ in the numerator which also arranges for the appropriate normalisation, see \cite{Lu:2023mkn}. The RG-invariance of $\bar G_q$ and hence that of the integrand  in \labelcref{eq:nq} is readily seen from 
\begin{align}
\tr\,\bigl[ \gamma_0\, \bar G_{q}(p_q)  \bigr]& = - 4 N_c \, \imag \, p_{q,0}\, Z_q^{E}(p_q)\, f_q(p_q,\mu_q)\,, 
	\label{eq:ZtrG}
\end{align}
with $f_q(p_q,\mu_q)$ in \labelcref{eq:ModeIntegrandsfqMain} and 
\begin{align} 
	Z_q^{E}(p_q)\, f_q(p_q,\mu_q) = \frac{1}{  p_{q,0}^2+\left(\frac{Z_q^M}{Z_q^E}\right)^2\boldsymbol{p}^2+M^2_q} \,.
	\label{eq:RGI-Zqfq} 
\end{align} 
The right hand side of \labelcref{eq:RGI-Zqfq} is manifestly RG-invariant and so are  \labelcref{eq:nq,eq:ZtrG}. The computational details of performing the thermal sums in \labelcref{eq:nq} can be found in \Cref{app:ComputationalDetails}. 

\begin{figure}[t]
	\centering
	\includegraphics[width=0.9\columnwidth]{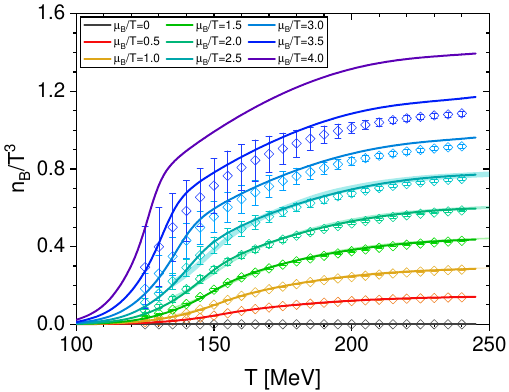}
	\caption{Baryon number density $n_B/T^3$ for different baryon chemical potential $\mu_B/T= 0,0.5,\cdots,4.0$ as a function of $T \in (100,250)\,\textrm{MeV}$. We also show lattice QCD data from the $T^{\prime}$ expansion scheme~\cite{Borsanyi:2021sxv} (open diamonds) and the sixth order Taylor expansion~\cite{Bollweg:2022rps} (coloured bands). }
	\label{fig:EoSTmuB}
\end{figure}

The baryon number density $n_B$ follows from that of the net-quark number densities as a flavour sum,  
\begin{align}
  n^{\ }_B(T,\mu_B) = \frac{1}{3} \sum_{q_i=u,d,s}  n_{q_i}(T,\mu_{q_i})\,. 
  \label{eq:nqtonB}
\end{align}
The temperature dependence of $n_B(T,\mu_B)$ including the confining background is displayed in \Cref{fig:EoSTmuB} for different baryon chemical potentials with $\mu_B/T$ from $0.5$ to $4.0$ in steps of $0.5$.  We also show extrapolation results from the lattice QCD, including those from the Taylor expansion up to $O(\mu_B/T)^6$~\cite{Bollweg:2022rps} (coloured bands), and those from the $T^{\prime}$ expansion scheme~\cite{Borsanyi:2021sxv} (open diamonds). 

Our result agrees well with the lattice QCD extrapolations for $\mu_B/T \leq 3.5$ within the error bars from the extrapolations. For $\mu_B/T>3.5$ we are beyond the convergence radius of lattice extrapolations. We also emphasise that our result of $n_B$ in \Cref{fig:EoSTmuB} are obtained with a direct functional computation without any external input. Hence, \Cref{fig:EoSTmuB} and further thermodynamic observables and the fluctuations of conserved charges at small density serve as decisive benchmarks for the present approach: to begin with, the present approach allows for direct computations at larger densities $\mu_B/T\gtrsim 3$ and hence, together with the fRG approach, is the only QCD-based approach at these densities that provides direct results. This leads to the following crucial observation:   
Assume that one obtains the  results agree with all lattice benchmark observables at small density,  and then extends to larger density with either direct lattice extrapolations or low energy effective theories such as NJL-type models, Quark-Meson models and holographic models.  Then, the results from the functional QCD approach at larger density have a \textit{qualitatively} better systematics than any of these extrapolations as it includes the QCD dynamics. This statement persists even in the regime with $\mu_B/T\gtrsim 4$, where the systematic error of functional QCD approaches still grows larger. Moreover, this deficiencies will be remedied soon by improved computations within the fQCD collaboration \cite{fQCD}. In any case, the present functional results for the crossover line including the location of the CEP and the results for the observables there should be considered as the benchmark for low energy effective model computations including lattice extrapolations. 

For small chemical potentials the results agree within the systematic error estimate of 10\% for the current results. For $\mu_B/T \lesssim 3$ the comparison is even far  better. For $\mu_B/T \gtrsim 3$ and large temperatures the $n_B$ from functional QCD grow slightly larger than those from the lattice extrapolation, while still within our systematic error estimate of 10\%. This overshooting tendency for larger temperatures was also present for the renormalised chiral condensate at vanishing chemical potential, see \Cref{fig:condA0}. Hence, it may hint at an overshooting of the integrated scalar dressing $Z^E_q \times f_q$, \labelcref{eq:RGI-Zqfq} within the systematic error. This overshooting is accommodated within the systematics of the difference DSE used in the current miniDSE scheme and is part of the 10\% systematic error estimate. It will be investigated and systematically improved in further works. 

Still, we note that the corresponding regime already overlaps with the predicted regime of novel phenomena, especially the moat regime which features spatial modulations in the thermodynamic system~\cite{Pisarski:2021qof,Fu:2024rto}. 
In other words, the density profile might contain signatures of the onset of novel phase structure of QCD at large chemical potentials. 
Such a property is most likely not captured by an extrapolation. 

We emphasise that the gluonic background $\varphi^c$ plays a crucial role in the thermodynamic behaviour of $n_B$ in the hadronic regime, in contradistinction to the irrelevance of the gluonic background for the chiral condensate shown in \Cref{fig:condA0}. 
We have deferred the respective discussion to \Cref{app:EoS}, see in particular \Cref{fig:chiB-thermo-comp}: the importance of the consistent $\varphi^c$-background is growing large for the hadronic regime.  
This illustrates very concisely the relevance of the confining dynamics in the finite-density part of QCD thermodynamic functions. 

In the present Section we concentrate on the physics results from the present approach, and proceed with the computation of the equation of state and further thermodynamic observables at finite density from the net-baryon number density. 
To that end we simply consider the density effects of the thermodynamic observables, and we explain the general procedure at the example of the equation of state: 
Key to the computation is the property, that the finite-density effects in the equation of state and other thermodynamic observables at a given $\mu_B$ are completely captured by a $\mu_B$-integral of the density~\cite{Lu:2023mkn, Lu:2023msn}. 

\begin{figure}[t!]
	\centering
	\includegraphics[width=0.9\columnwidth]{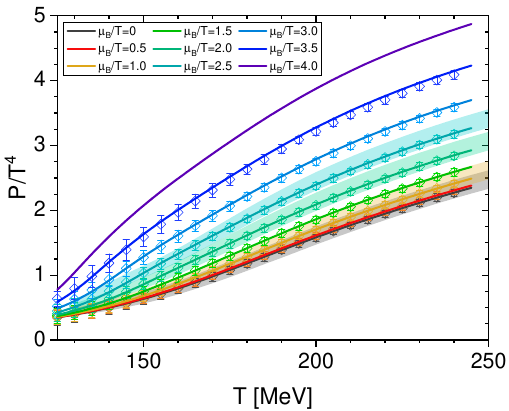}\\[2ex]
	\includegraphics[width=0.9\columnwidth]{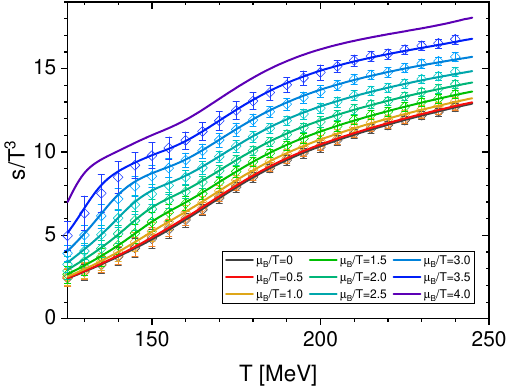}\\[2ex]
	\includegraphics[width=0.9\columnwidth]{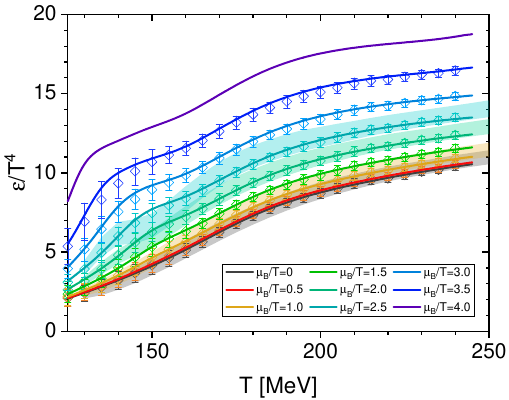}
	\caption{QCD equations of state at finite temperature and density: pressure ($P/T^4$), entropy density ($s/T^3$) and energy density ($\epsilon/T^4$) as  functions of temperature with $T \in (100,250)\,\textrm{MeV}$ for baryon chemical potentials $\mu_B/T = 0,0.5,\cdots,4.0$. We also show lattice QCD data from the $T^{\prime}$ expansion scheme~\cite{Borsanyi:2021sxv} (open diamonds) and the Taylor expansion~\cite{Bazavov:2017dus} (coloured bands). }
	\label{fig:EoST+Entropy+EnergyDensitymuB}
\end{figure}
For its importance for the present results and in order to allow for a direct systematic error assessment we illustrate the approach at the example of the pressure. It reads 
\begin{align}
	P(T,\mu_B) = P(T,0) + \int\limits_0^{\mu_B} \mathrm{d}\mu\, n^{\ }_B(T,\mu) \,, 
	\label{eq:EoSchem}
\end{align}
which can be computed from $P(T,0)$ and from our density results \Cref{fig:EoSTmuB}. The entropy follows as its temperature derivative,  
\begin{align}
	s(T,\mu_B) = \frac{ \partial P(T,\mu_B) }{ \partial T }\,.  
\label{eq:Entropy}
\end{align}
Hence, using \labelcref{eq:EoSchem}, the entropy at finite baryon chemical potential can be computed from the input $s(T,0)$ and $\partial/\partial  n_B(T,\mu_B)$, computed in the present work. Finally, these results for the pressure $P(T,\mu_B)$ and the entropy $s(T,\mu_B)$ can be used directly to determine the energy density 
\begin{align}
\hspace{-.1cm}	\epsilon(T,\mu_B) =T s(T,\mu_B) - P(T,\mu_B)+ \mu_B \,n_B(T,\mu_B) \,,
	\label{eq:EnergyDenstiy} 
\end{align}
and the trace anomaly 
\begin{align}
 I(T,\mu_B) & = \left[ \epsilon(T,\mu_B) - 3P(T,\mu_B) \right] /T^4 \notag \\
 & = T \frac{\partial}{\partial T} \left( \frac{P(T,\mu_B)}{T^4} \right) + \frac{\mu_B}{T^4} \, \frac{\partial P(T,\mu_B)}{\partial \mu_B}\,.
\label{eq:TA0}
\end{align} 
At vanishing baryon chemical potential $\mu_B$ the thermodynamic observables, and in particular $P, s, \epsilon, I$ have been determined very accurately with lattice simulations and we use the $\mu_B=0$ results from~\cite{Borsanyi:2021sxv}. 
\begin{figure}[t!]
	\centering
	\includegraphics[width=0.9\columnwidth]{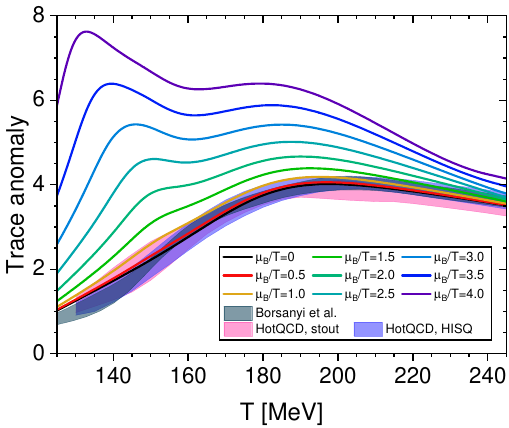} 
	\caption{Trace anomaly $I=(\epsilon-3P)/T^4$ as a function of temperature at given $\mu_B/T = 0,0.5,\cdots,4.0$. We also show lattice QCD data~\cite{HotQCD:2014kol} (hotQCD) and~\cite{Borsanyi:2025dyp} (Borsanyi et al.) for $\mu_B=0$. }
	\label{fig:TraceAnomalymuB}
\end{figure}
\begin{figure*}[t]
	\centering
	\includegraphics[width=0.95\columnwidth]{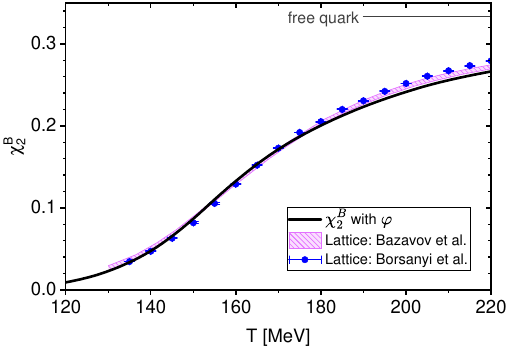} 
	\hspace{.2cm}
	\includegraphics[width=0.9735\columnwidth]{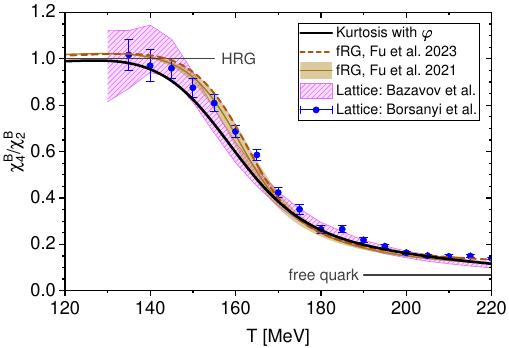}
	\caption{Baryon number susceptibilities $\chi_2^{B}$ and the kurtosis $\chi_4^{B}/\chi_2^{B}$ as a function of temperature at $\mu_B = 0$ and $\mu_Q=\mu_S=0$. 
		The lattice QCD results from \cite{Bazavov:2017dus,Borsanyi:2018grb}, and the fRG-LEFT results from \cite{Fu:2021oaw, Fu:2023lcm}, are shown together for comparison. }
	\label{fig:chiB-zeromu}
\end{figure*}

Our results for the QCD equations of state, the pressure $P$, entropy density $s$ and the energy density $\epsilon$, are shown in \Cref{fig:EoST+Entropy+EnergyDensitymuB} together with the lattice data from \cite{Borsanyi:2021sxv} and \cite{Bazavov:2017dus}. We display the results for $\mu_B/T$ ranging from 0 to 4.0 in 0.5 steps. These plots are complemented by that on the trace anomaly in \Cref{fig:TraceAnomalymuB}. 

The results in \Cref{fig:EoST+Entropy+EnergyDensitymuB} may be seen as a benchmark test for the accuracy of the finite density correlations (for the quark propagator) in the present functional QCD approach. However, they should rather been seen as a confirmation of the validity of lattice QCD extrapolations of the equation of state at small chemical potentials. Note that the latter is done with only a few parameters. 

We have already mentioned that the impressive agreement extends to the 
baryon number susceptibilities $\chi_2^B$ and $\chi_4^B$, see \Cref{sec:baryon-flucs}. In contradistinction to the thermodynamic observables the results for the susceptibilities is computed solely from the correlation function results of the functional QCD approach used here. 

Finally we also comment on the $\mu_B/T = 3.5$ results in comparison to the lattice extrapolations. This is already close or even beyond the convergence radius of the lattice QCD extrapolation. Consequently, the error bars of the lattice results are growing large, particularly in the low temperature region.  

In summary, these results confirm the quantitative accuracy of the present DSE computations for $\mu_B/T\lesssim 3$. As discussed before, the results for $P,s,\epsilon, I$ for $\mu_B/T\lesssim 3$ constitute the first fully self-consistent functional QCD computation of these observables in this regime. In the absence of any other first principles QCD result, these are the first QCD results in this regime. Hence, they set the benchmark for the thermodynamic observables in this regime including the critical end point. Extrapolation results and that from low energy effective theories (LEFT) should be confronted with them and potentially adjusted. 

The present results provide a direct access on the equation of state at relatively high densities, which is of imminent importance for the theoretical access to experimental data from low-energy collision experiments. 
They can be used as input for respective  phenomenological studies, e.g.~ in hydrodynamic simulations.
Related investigations are beyond the scope of the present work and are subject of  forthcoming ones.

%%%%%%%%%%%%%%%%%%%%%%%%%%%%%%%%%%%
\subsection{Fluctuations of the conserved baryon charge}
\label{sec:baryon-flucs}

We have computed thermodynamic observables (the equations of state) of QCD at finite baryon chemical potential in \Cref{sec:nB+Thermodynamics}. 
This has been done by representing their density contributions via the baryon density, and its temperature integral  (pressure) and first order derivatives with respective to temperature (entropy, energy density) or baryon chemical potential (trace anomaly). 
Higher order $T,\mu_B$-derivatives provide a differential resolution of the equations of states, and hence provide more information about the underlying QCD dynamics.  
The $n$th order derivatives with respect to baryon chemical potential of the pressure are the baryon number susceptibilities $\chi_{n}^{B}$ with 
\begin{align}
  \chi_{n}^{B} = \frac{\partial^n (P/T^4)}{\partial(\mu_B/T)^n} = T^{n-4} \frac{\partial^{n-1} n_B}{\partial \mu_B^{n-1}}\,.
\label{eq:chiBn}
\end{align}
The $ \chi_{n}^{B}$ give access to a whole set of observables, and in particular to the baryon number fluctuations. The latter are closely related to the fluctuations of net-proton number, measured in the experiments. 

\Cref{eq:chiBn} can be either computed directly on the numerical results for the baryon number density $n_B$ in \labelcref{eq:nq}. 
This requires a rapidly increasing numerical precision for $n_B$ with rising order $n$ of derivatives. 
A more promising alternative is taking analytic $\mu_B$ derivatives of \labelcref{eq:nq}, hitting the correlation functions as well as the explicit $\mu_B$-dependence, see \cite{Fu:2015naa, Faigle-Cedzich:2023rxd}. 
There it has been shown that the higher order $\mu_B$-derivatives of correlation functions or rather the dressings can be iteratively computed from the lower ones without the need of taking numerical derivatives. 
For example, $\chi_2^B$ is related to the $\mu_B$-derivative of the quark number density, 
\begin{align}\nonumber 
  \frac{\partial n_q}{\partial \mu_B} =& \, - \sumint_{p} \, \tr \Biggl[ \gamma_0 \Biggl(\frac{\partial \bar{G}_{q} (p)}{\partial \mu_B} \\[1ex]  
  &\hspace{2.3cm}+ \sum_{i} \frac{\partial \mathcal{D}_i (p)}{\partial \mu_B}\frac{\partial \bar{G}_{q}}{\partial D_i}(p)  \Biggr) \Biggr]\,,
\label{eq:dnqdmu}
\end{align}
where $\bar G_q$ is the RG-invariant normalised quark propagator \labelcref{eq:barGq}. The $\mathcal{D}_i$'s are parts of correlation functions such as wave functions or mass functions as well as dynamical backgrounds such as the dynamical background $\varphi^c$ which signals confinement. In the present investigation we have to consider the quark dressing functions and the  $\varphi^c$,  
\begin{align}
\mathcal{D} = \{Z_q^E\,,\,Z_q^M\,,\,M_q\,,\,\varphi^c\}\,,
\label{eq:setD} 
\end{align} 
Similarly, higher order baryon number susceptibilities can be obtained by further applying the $\mu_B$ derivatives on \labelcref{eq:dnqdmu} order by order. In the present work we have used the first step in this iteration and computed the numerical derivatives for the ${\cal D}_i$ instead of using the full iterative procedure. Close to the crossover line this led to the restriction $\mu_B\lesssim 600$\,MeV for the results, the larger $\mu_B$ regime will be discussed elsewhere. On the freeze-out line discussed later, the numerical accuracy is sufficient to go to larger $\mu_B$, that is smaller $\sqrt{s_{NN}}$, see \Cref{fig:R42-Frz}. 

\begin{figure}[b]
	\centering
	\includegraphics[width=0.9\columnwidth]{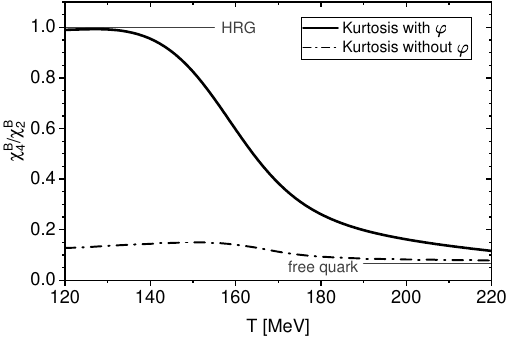} 
	\caption{Baryon number kurtosis $\chi_4^{B}/\chi_2^{B}$ at vanishing $\mu_B$, evaluated with the dynamical gluonic background $\varphi$ and at $\varphi=0$. Computations in the latter background are not converged and fail to even meet the qualitative features, including $\chi_4^{B}/\chi_2^{B}=1$ at $T=0$.}
	\label{fig:R42-comp-main}
\end{figure}

In \Cref{fig:chiB-zeromu}, we show our results for the second-order baryon number susceptibility $\chi_2^B$ and the kurtosis $\chi_4^B/\chi_2^B$ at $\mu_B=0$. 
The computation draws from the complete sets of correlation functions and potentials computed here and we provide a brief review of the most important properties: 
 
To begin with, they are obtained in the dynamical gluonic background $\varphi$: we have discussed in \Cref{sec:Fun+Conf,sec:Conf-Deconf} and \Cref{app:Poloop+PolPot} that this background signals the transition from the  quark-gluon phase at large temperatures to the confined hadronic phase at low temperatures. 
Its inclusion is of crucial importance for even obtaining the qualitative properties of the fluctuations of conserved charges, see \cite{Fu:2015naa, Fu:2016tey}. We have repeated the respective analysis here. 
\Cref{fig:R42-comp-main} comprises the comparison of the kurtosis with and without the gluonic background $\varphi$. This comparison illustrates the qualitative importance of the dynamical gluonic background, that signals confinement. 
Specifically, within the dynamical gluonic background we meet the crucial benchmark for the kurtosis for vanishing temperature and density, $ \chi_4^B/\chi_2^B(T=0)=1$. 
This signals the baryonic nature of the hadronic phase, while in the quark regime is signalled by  $\chi_4^B/\chi_2^B\to 1/9$. 
Monitoring this dynamical change accurately of degrees of freedom is vital for the explanation of the dynamics in the crossover regime that shows in the experimental data for fluctuations of conserved charges. 
This important property as well as the underlying dynamics is not even captured qualitatively for vanishing gluonic background $\varphi=0$. A more detailed comparison, and in particular that of the importance of $\varphi_8 \in \imag \,\mathbbm{R}$ is done in \Cref{app:EoS}. 

On the more technical side we remark that the computation of observables within the dynamical gluonic background $\varphi$ optimised the convergence of the computation as $\varphi$ is the solution of its equation of motion. 
We conclude that the present results constitute the first functional QCD results for the fluctuations of conserved charges that accommodates all qualitative properties of the QCD dynamics in the crossover regime and in the hadronic phase as well as providing quantitative results. 
The results are obtained within a self-contained functional computation: all correlation functions are computed within the present DSE approach and the gluonic background is obtained from the functional Polyakov loop potential obtained within the present DSE approach, see \Cref{sec:Conf-Deconf} and \Cref{app:Poloop+PolPot}. This concludes the brief overview. 

We proceed with the discussion of our results for the kurtosis. In \Cref{fig:chiB-zeromu} we also show benchmark results from lattice QCD \cite{Bazavov:2017dus, Borsanyi:2018grb} and from QCD-assisted fRG-LEFTs \cite{Fu:2021oaw, Fu:2023lcm}. 
Our result agree quantitatively with these benchmark results for all temperatures considered. 

\begin{figure}[t]
	\centering
	\includegraphics[width=0.89\columnwidth]{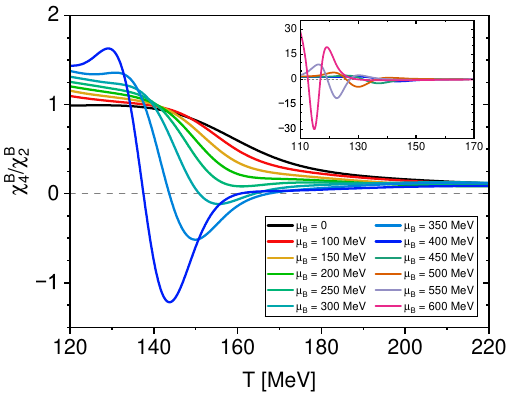}
	\caption{Baryon number kurtosis $\chi_4^{B}/\chi_2^{B}$ as a function of temperature for different baryon chemical potentials ranging from $\mu_B=0$ to 600\,MeV in steps of 50\,MeV.   }
	\label{fig:kurtosis-finite-muB}
\end{figure}
With this preparations we proceed to QCD at finite baryon chemical potential, and in particular to $\mu_B/T\gtrsim 3$. There, the present approach gives us access to a precision analysis of thermodynamic observables at finite density, particularly for the event-by-event fluctuations of conversed charges that are directly accessible in heavy-ion collisions. 
The respective results for the kurtosis at finite $\mu_B$, are collected in \Cref{fig:kurtosis-finite-muB}. There we show the temperature dependence of the kurtosis for $\mu_B$, ranging from 0 to 600\,MeV. 
In particular we find that the non-monotonicity of the kurtosis becomes more sizeable with increasing $\mu_B$. 
This is expected from previous studies both in QCD and low energy effective theories. 
Moreover, it follows directly from the basic properties of analytic functions. While such a behaviour could signal critical phenomena for the fluctuations observed in the low-energy collisions, it has been argued that the critical regime in QCD is very small and no functional QCD computation shows critical scaling for $\mu_B\lesssim 400$\,MeV, for a recent discussion see \cite{Fu:2023lcm}. 
Consequently we consider it as very unlikely that the non-monotonicities signal critical scaling. Indeed, for a dissection of the QCD phase structure and in particular the location of the critical end point the non-critical scenario is far more amiable towards its detection, see again \cite{Fu:2023lcm}. 

We close this Section with a brief discussion of the systematics: the current approximation is quantitatively accurate for $\mu_B/T\lesssim 4$ and the quantitative agreement of the chiral crossover lines from functional QCD, \cite{Fu:2019hdw, Gao:2020fbl, Gunkel:2021oya} for $\mu_B/T\gtrsim 4$ provides non-trivial support for its predictive power for larger baryon chemical potentials including the CEP regime. However, this argument does not apply to the susceptibilities and further differential observables, and the systematic error of the respective results grows successively larger for $\mu_B/T\gtrsim 4$.

%%%%%%%%%%%%%%%%%%%%%%%%%%%%%%%%%%%%%
\subsection{Baryon number kurtosis and beam energy scan}
\label{sec:eos-A0-bes}

\begin{figure}[t]
	\centering
	\includegraphics[width=1.0\columnwidth]{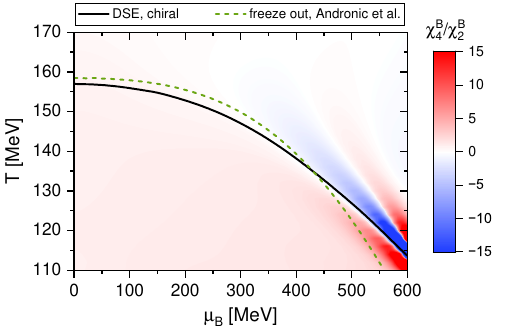}
	\caption{Kurtosis $\chi_4^B/\chi_2^B$ as a function of temperature and baryon chemical potential $\mu_B$. We also display the 
		the chiral crossover line and the freeze-out line extracted with the statistical hadronisation approach~\cite{Andronic:2017pug} as a frame of reference.
	}
	\label{fig:R42-muB-2D}
\end{figure}
In this final part of \Cref{sec:ResultsObservables} we map out the kurtosis in the phase structure, concentrating on the crossover regime and the approach to the critical end point (CEP), see also \cite{Fu:2023lcm}. For a recent review concerning its relevance for detecting the CEP, see \cite{Stephanov:2024xkn}. This result is used for analysing the behaviour of the baryon number kurtosis on the freeze-out line with respect to possible signatures in the beam energy scan of heavy-ion collisions. The experimental measurements provide the net-proton cumulants, and hence this comparison has to be taken with a grain of salt. The theoretical equilibrium results should be seen as the QCD fundament for going towards the experimental signatures, and these additional steps not only have to accommodate the difference of the fluctuations of conserved charged but also have to take care of the non-equilibrium nature of the heavy ion collision. Still, the QCD fundament is most important for anchoring any progress towards the experimental signatures. 

\begin{figure}[t]
	\centering
	\includegraphics[width=0.86\columnwidth]{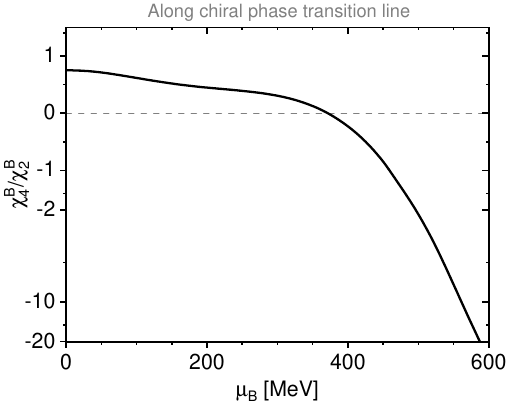}
	\caption{Kurtosis $\chi_4^B/\chi_2^B$ along the chiral phase transition $T_\chi(\mu_B)$ line as a function of $\mu_B$. }
	\label{fig:R42-chiralPT}
\end{figure}

In \Cref{fig:R42-muB-2D} we show a heat map plot of $\chi_4^B/\chi_2^B$ up to baryon chemical potentials $\mu_B = 600$\,MeV. 
This value is already close to the location \labelcref{eq:LocationCEP} of the CEP in the present computation with $\mu^\textrm{CEP} _B=660$\,MeV. The heat map \Cref{fig:R42-muB-2D} is complemented with the kurtosis along the chiral crossover line in \Cref{fig:R42-chiralPT}. 
The result is depicted on a linear scale for $|\chi_4^B/\chi_2^B| < 1$ and a logarithmic scale for $|\chi_4^B/\chi_2^B| > 1$. 
We find that the kurtosis starts out positive and turns negative at $\mu_B$ between $350 - 400$\,MeV.
 This behaviour has also been seen in the fRG studies within QCD-assisted LEFTs, \cite{Fu:2021oaw, Fu:2023lcm}. 
 It is a common property of generic low energy effective theories, see e.g.~\cite{Fu:2016tey}. 
 Moreover, in such a setup it is readily shown that it persists in the absence of a critical end point. 
 In this context it is also worth emphasising that critical scaling in this regime is neither present in functional QCD studies \cite{Fu:2019hdw, Gao:2020fbl, Gunkel:2021oya} nor in the low energy effective theory ones. 
 In conclusion, while such non-monotonicities do occur in the vicinity of a critical scaling regime around the CEP, they are no smoking gun for it. 
 Indeed, in this regime we still are non-critical which supports a non-critical search for the critical end point. 
 Nevertheless, it is a huge advantage that such a search does not rely on the extraction of critical properties from relatively noisy experimental data, for a detailed discussion of this important point see \cite{Fu:2023lcm}.  

With these preparations we also investigate the kurtosis along the freeze-out line. This provides the first  
equilibrium baseline for the beam energy scan experiment from direct functional QCD computations. Moreover, in the absence of lattice QCD results for collision energies $\sqrt{s_{NN}}\lesssim 8$\,GeV, the present results are the first QCD results available. 
For the freeze-out line we use the parametrisation~\cite{Andronic:2017pug}, based on  the statistical hadronisation approach. 

\begin{figure}[t]
	\centering
	\includegraphics[width=0.86\columnwidth]{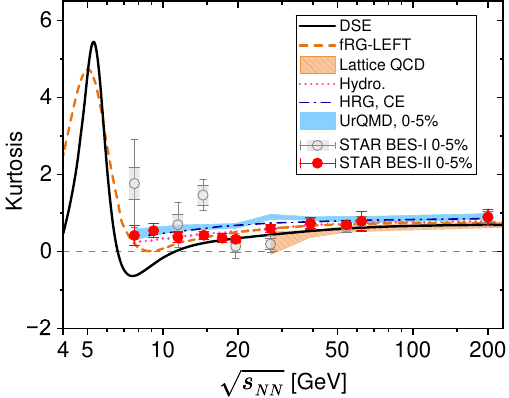}
	\caption{Kurtosis $\chi_4^B/\chi_2^B$ along the freeze-out line~\cite{Andronic:2017pug} as a function of collision energy $\sqrt{s_{NN}}$. We also show results with a canonical ensemble within the QCD-assisted fRG-LEFT~\cite{Fu:2023lcm}, the lattice QCD result~\cite{Bazavov:2020bjn}, the STAR BES-I~\cite{STAR:2020tga} and BES-II~\cite{STAR:2025zdq} data for the net-proton cumulant ratio $C_4/C_2$, together with baselines from HRG~\cite{Braun-Munzinger:2020jbk}, UrQMD~\cite{STAR:2021iop} and hydrodynamic simulations~\cite{Vovchenko:2021kxx}. }
		\label{fig:R42-Frz}
\end{figure}

In \Cref{fig:R42-Frz} we show the kurtosis of net baryon number as a function of $\sqrt{s_{NN}}$ along the freeze-out line from ~\cite{Andronic:2017pug}. The results are complemented  by the fRG data from \cite{Fu:2023lcm}, using corrections for the canonical ensemble. This work is done in a sophisticated QCD-assisted low energy effective theory (LEFT) that features the chiral physics and crossover line obtained in functional QCD in \cite{Fu:2019hdw}. However, the setup still allowed for minor variations of the location of the CEP with approximately $\pm 50$\,MeV relative to the present one in \labelcref{eq:LocationCEP}. Such a variation led to large variations of the amplitude of the cumulants on the freeze-out line for $\sqrt{s_{NN}}\lesssim 10$\,GeV. In turn, the peak location of the kurtosis on the freeze-out line only showed subleading variations, see Figure 11 in \cite{Fu:2023lcm}. The respective analysis readily translates to the present setup: the amplitude of the kurtosis at the freeze-out line shown in \Cref{fig:R42-Frz} may change with an improved computation as the location of the CEP still has a (conservative) systematic error of $\pm 50$\,MeV. In turn, the peak position can be already read-off from \Cref{fig:R42-Frz} with a good accuracy, and matches the one of \cite{Fu:2023lcm}. Note that this analysis is true for both the canonical and grand canonical ensemble. 

The plot also contains the STAR BES-I~\cite{STAR:2020tga} and BES-II~\cite{STAR:2025zdq} data of the kurtosis of net-proton number, the lattice QCD calculation from \cite{Bazavov:2020bjn}, as well as the baselines obtained from HRG \cite{Braun-Munzinger:2020jbk}, UrQMD \cite{STAR:2021iop}, and the hydrodynamic simulation \cite{Vovchenko:2021kxx}. 
The lowest energy point of BES-II is $7.7\,$GeV, and for lower energies future experiments such as CEE+ (HIAF),  CBM (FAIR) and NICA (JINR) are taking data soon. 

%

%%%%%%%%%%%%%%%%%%%%%%%
\section{summary}
\label{sec:Summary}

We have computed thermodynamic observables and fluctuations of conserved charges within first principles functional QCD. The computational Dyson-Schwinger setup draws from earlier investigations and technical developments \cite{Gao:2020qsj, Gao:2020fbl,Lu:2023mkn}. The crucial novel ingredient for the computation of these observables is the use of the dynamical gluonic background $\varphi^c$ that signals confinement, see \Cref{sec:Conf-Deconf} and  \Cref{app:Poloop+PolPot}. This background constitutes a solution of the gluonic equations of motion and carries a considerable part of the confining infrared dynamics of QCD. It is gauge invariant and is related to both the temporal component of the gauge field and the Polyakov loop. On the technical side we note that this background can be seen as an optimal and stable expansion point, and hence ensures a rapid convergence of the computation. Notably, for $\varphi^c=0$ the above mentioned observables lack even qualitative features in the hadronic regime and cannot be used to access the transition regime around the chiral and confinement-deconfinement crossovers. In turn, the results within the gluonic background are quantitatively trustworthy.   

We find that the confinement-deconfinement crossover is close to the chiral one but features a far larger width, see \Cref{fig:PD-conf-width}. Both crossovers end in the (chiral) critical end point, see \Cref{fig:PhaseStructureChiralConf2025}. The thermodynamic observables, that is the pressure, entropy, energy density and trace anomaly, are in quantitative agreement with the lattice benchmarks at vanishing and small densities with $\mu_B/T\lesssim 3$. For larger densities, the present results constitute the first functional QCD without any external input, and constitute the QCD benchmark for $\mu_B/T\gtrsim 3$, see \Cref{fig:EoST+Entropy+EnergyDensitymuB,fig:TraceAnomalymuB}. 

We have also studies baryon number susceptibilities, and computed the 2nd and 4th order cumulants. As for the other observables, they meet the lattice benchmarks well, see \Cref{fig:chiB-zeromu} and constitute the QCD benchmarks for $\mu_B/T\gtrsim 3$, see \Cref{fig:kurtosis-finite-muB}. This allows us to map out the kurtosis in the QCD phase structure, the respective heat map is provided in \Cref{fig:R42-muB-2D}. Moreover, we have used the results to follow the kurtosis on the chiral crossover line, \Cref{fig:R42-chiralPT} as well as on the freeze-out line, \Cref{fig:R42-Frz}. There, our results constitute the equilibrium QCD baseline, in particular for $\sqrt{s_{NN}}\lesssim 8$\,GeV, in the energy regime of the future heavy ion experiments CEE+ (HIAF), CBM (FAIR), NICA (JINR), taking data soon, and hopefully also NA60+, NA61/SHINE (CERN). There, this baseline has to be confronted with the event-by-event measurement of the baryon number fluctuations in these experiments. Note however, that apart from the equilibrium nature of the present computation, further differences have to be considered as baryon number versus proton number fluctuations and canonical versus grand canonical ensembles. This and further also computational improvements relevant at higher densities are under completion, and we hope to report on them soon.

%%%%%%%%%%%%%%%%%%%%%%%
\begin{acknowledgments}

The authors thank Jens Braun, Maurício Ferreira, Wei-jie Fu, Chuang Huang, Joannis Papavassiliou, Fabian Rennecke, Franz Sattler, Shi Yin and Rui Wen for discussions. This work is done within the fQCD collaboration \cite{fQCD}, and we also thank the other members of the collaboration for discussions and collaborations on related subjects. This work is funded by the Deutsche Forschungsgemeinschaft (DFG, German Research Foundation) under Germany's Excellence Strategy EXC 2181/1 - 390900948 (the Heidelberg STRUCTURES Excellence Cluster) and the Collaborative Research Centre SFB 1225 - 273811115 (ISOQUANT). JMP acknowledges support by the Chinese Academy of Sciences President's International Fellowship Initiative Grant No.~2024PG0023. YL and YXL are supported by the National  Science Foundation of China under Grants  No.~12175007 and No. ~12247107.  FG is supported by the National  Science Foundation of China under Grants  No.~12305134. 

\end{acknowledgments}

\appendix

%%%%%%%%%%%%%%%%%%%%%%%%%%%%%%%%
\section{Center-symmetry breaking in finite density QCD}
\label{app:Poloop+PolPot}

In this Appendix we discuss the functional approach to center-symmetry breaking in QCD at finite density. This approach has been set up in the functional renormalisation group, including results to the QCD phase structure in \cite{Braun:2007bx, Braun:2009gm, Braun:2010cy, Marhauser:2008fz, Fister:2013bh, Herbst:2015ona}. Its setup for DSE has been introduced in \cite{Fischer:2009wc, Fister:2013bh}, and applications can be found in \cite{Fister:2013bh, Fischer:2013eca, Fischer:2014ata, Fischer:2014vxa}, for fRG-review also discussing some of the subtleties see \cite{Dupuis:2020fhh, Fu:2022gou}. Furthermore, it has been also picked up and developed further in the Curci-Ferrari model~
\cite{Reinosa:2014ooa, Reinosa:2015oua, Maelger:2017amh, 10.21468/SciPostPhys.12.3.087, vanEgmond:2024ljf, MariSurkau:2025pfl}, see specifically~\cite{10.21468/SciPostPhys.12.3.087, vanEgmond:2024ljf} for a manifestly center-symmetric formulation. 

While for dynamical quarks the relation of center-symmetry breaking to the confinement-deconfinement phase transition is less clear, the respective non-trivial temporal background field $A_0$, related to the expectation value of the Polyakov loop, is essential for the rapid convergence of expansion schemes in functional approaches, for a discussion in pure Yang-Mills theory see \cite{Cyrol:2017qkl}. Specifically, the $A_0$-background is a crucial ingredient for the computation of the thermodynamic observables evaluated in the present work. While this property of the $A_0$-background or the respective Polyakov loop has been called `statistical confinement', we consider this a misnomer for the reasons above: it simply carries a crucial dynamical information similar to that carries by the chiral condensate in the context of dynamical chiral symmetry breaking. 

For its importance for the present work we present a rather detailed introduction to this approach, including also computational instructions. While most of it can be found in the references cited above, we believe, that the present introduction helps the reader both for their understanding but also for own computations. In \Cref{app:PloopAlg} we introduce the gauge-invariant functional order parameter for center-symmetry breaking suggested in \cite{Braun:2007bx} and discuss its relation to the $A_0$-background as well as the expectation value of the traced Polyakov loop. In \Cref{app:DSE-Polya} we provide a computational guideline for the computation of the related order parameter potential within the DSE, including the subtleties at finite density.

%%%%%%%%%%%%%%%%%%%%%%%%%%%%%%%%
\subsection{Order parameter for center-symmetry breaking in functional approaches}
\label{app:PloopAlg}

In this Appendix we provide a brief review of the functional approach to center-symmetry breaking put forward in \cite{Braun:2007bx, Braun:2010cy, Marhauser:2008fz, Fister:2013bh, Herbst:2015ona}. We restrict ourselves to the physical case of SU(3) important for the present work.

%%%%%%%%%%%%%%%%%%%%%%%%%
\subsubsection{Traced Polyakov loop} 
\label{app:PolLoop}

The traced Polyakov loop in the fundamental representation is sensitive to center symmetry transformation. It may be understood as the interacting part of a static quark, sitting at the position $\boldsymbol{x}$ for all times. Such a quark solves the static Dirac equation for all times $t$, given the quark at some initial time $t_0$, 
\begin{align} 
	\slashed{D} \,q(x) = 0\,,\quad q(x) = P(t,t';\boldsymbol{x})\, q(x')\,,
\end{align} 
with a vanishing spatial gauge field $\boldsymbol{A}=0$ and $x= (t,\boldsymbol{x})$, $x'= (t',\boldsymbol{x})$. The temporal Wilson line is given by 
\begin{align} 
	P(t^\prime,t;\boldsymbol{x})= \mathcal{P}\, e^{ -\imag g_s \int\limits_{t'}^t d \tau \,A_0(\tau,\boldsymbol{x}) }\,.
\label{eq:P+barP}
\end{align}
In \labelcref{eq:P+barP}, ${\cal P}$ is the path ordering and $g_s$ is the strong coupling and $\boldsymbol{x}$ is the spatial position. We now consider static quarks and anti-quarks at finite temperature with the initial time $t^\prime=0$ and the final time $t=\beta$. Then the phase factor is given  
\begin{align} 
 P(\boldsymbol{x})= P(0,\beta;\boldsymbol{x})\,.
	\label{eq:P}
\end{align}
Now we consider the free energy of a quark--anti-quark pair at the positions $\boldsymbol{x}$ and $\boldsymbol{y}$ for asymptotic large distances, 
\begin{align}\nonumber 
	\lim_{r\to \infty} \langle \bar q(\boldsymbol{x}) q(\boldsymbol{y})\rangle \propto &\,	\lim_{r\to \infty} \langle \tr_f P(\boldsymbol{x}) \, \tr_f \bar P(\boldsymbol{y})\rangle \\[1ex]
	=& \langle \tr_f P(\boldsymbol{x})\rangle \langle   \tr_f \bar P(\boldsymbol{y})\rangle\,, 
	\label{eq:qbarqFreeEnegry} 
\end{align}
with $r=\|\boldsymbol{x} - \boldsymbol{y} \|$ and  
\begin{align}
	\bar P(\boldsymbol{x}) =  \mathcal{P}\, e^{ \imag g_s \int\limits_0^\beta d \tau \,A_0(\tau,\boldsymbol{x}) }\,.
	\label{eq:barP}
\end{align}
In the first step in \labelcref{eq:qbarqFreeEnegry} we dropped the free quark part of the expectation value, and in the second step we have used the cluster decomposition property. Moreover, for real gauge fields, \labelcref{eq:barP} is simply the adjoint of $P(\boldsymbol{x})$. The free energy $F_{q\bar q}$ of the quark--anti-quark state is proportional to minus the logarithm of \labelcref{eq:qbarqFreeEnegry} and for large distances it is proportional to 
\begin{align}
\lim_{r\to \infty} F_{q\bar q}(r) \propto -\left|\log \langle L\rangle +  \log \langle \bar L\rangle \right|\,,
\end{align}
with the normalised traced Polyakov loops     
\begin{align}
	{L}(\boldsymbol{x})= \frac{1}{N_c} \tr_\textrm{f}\, P(\boldsymbol{x}) \,,\qquad 	\bar {L}(\boldsymbol{x})= \frac{1}{N_c} \tr_\textrm{f}\, \bar P(\boldsymbol{x}) \,.
	\label{eq:TracedPLoop}
\end{align} 
Accordingly, in the confining phase of Yang-Mills theory with a diverging free energy of a quark--anti-quark the expectation value of the Polyakov loop has to be vanishing. In pure Yang-Mills theory the confinement-deconfinement phase transition is phase transition between a disordered center symmetric phase at low temperatures and a center-broken one at high temperatures: a center transformation with $z\in {\cal Z}_3$ leads to 
\begin{subequations}
	\label{eq:CenterTrafos} 
\begin{align} 
	P\to z\,P\,,\quad z=e^{2 \pi \imag \theta_z}\,,
	\label{eq:CenterTrafos1} 
\end{align}
with 
\begin{align} 
 \theta_z \in \left\{0\,,\, \frac{2}{\sqrt{3}} t^8 \,,\, t^3 - \frac{1}{\sqrt{3}} t^8 \right\}, 
	\label{eq:CenterTrafos2} 
\end{align}
\end{subequations}
The $t^a= \lambda^a/2$ are the SU(3)-generators with the Gell-Mann matrices $\lambda^a$ with $a=1,...,8$. With \labelcref{eq:CenterTrafos}, the traced Polyakov loop in the fundamental representation, $L$ in \labelcref{eq:TracedPLoop}, acquires a center phase under a center transformation. Consequently, its expectation value serves as an order parameter for center-symmetry breaking, and has the benefit that it is easily accessibly within lattice simulations.

%%%%%%%%%%%%%%%%%%%%%%%%%
\subsubsection{Gauge invariant eigenvalue field of the Polyakov loop} 
\label{app:Eigenvaluefield}

In terms of the correlation functions of gauge fixed QCD computed in the present work, $\langle L\rangle $ is obtained as a sum over $A_0$-correlation functions of any order. Apart from the fact that such an expectation value is difficult to access within functional approaches, in our opinion a better-suited order parameter is provided by the eigenvalue fields $ \nu_i $, that is derived from the algebra element $\varphi(\boldsymbol{x})$ of $P(\boldsymbol{x})$. The algebra field and its normalisation is defined with 
\begin{align}
	P(\boldsymbol{x}) = e^{2 \pi\imag \,\varphi(\boldsymbol{x})}\,,
	\label{eq:AlgebraField}
\end{align}
and the eigenvalues and eigenvectors in the fundamental representation follows as 
\begin{align}
	\varphi(\boldsymbol{x})\left|\psi_{\nu_i}\right\rangle = \nu_i(\boldsymbol{x}) \left|\psi_{\nu_i}\right\rangle \,, \qquad   i=1,2,3\,.
	\label{eq:Eigenvalues+vectors}
\end{align}
The Polyakov loop $P(\boldsymbol{x})$ transforms as a tensor under gauge transformations $U(x)\in$SU(3), and so does the algebra field $\varphi(\boldsymbol{x})$.  Accordingly, the eigenvalue fields $ \nu_i (\boldsymbol{x}) $ are gauge-invariant as the eigenvalues are invariant under unitary rotations such as gauge transformation. This can be used to write the algebra field as 
\begin{align}
	\varphi(\boldsymbol{x})= U^\dagger_\varphi(0,\boldsymbol{x})\,\nu(\boldsymbol{x})\,U_\varphi(0,\boldsymbol{x})\, , 
	\label{eq:nuvarphiGaugeTrafo}
\end{align}
where $\nu(\boldsymbol{x})$ is the diagonal eigenvalue field and $U_\varphi$ is the gauge transformation, that rotates $\varphi(\boldsymbol{x})$ into the Cartan subalgebra. Applying this rotation on the level of the gauge field we find  
\begin{align} 
\varphi^c(\boldsymbol{x}) = \frac{g_s\beta A^c_0}{2 \pi}\,, 
\label{eq:nuA}
\end{align} 
where the superscript ${}^c$ indicates that both, $\varphi^c=\nu$ and $A_0$, are in the Cartan subalgebra. 

It is left to construct an order parameter of center-symmetry breaking from the algebra field or rather the eigenvalue field. To that end we consider the center transformations of these fields, that follow readily from that of the Polyakov loop in \labelcref{eq:CenterTrafos}. This leads us to 
\begin{align} 
	\varphi(\boldsymbol{x}) \to 	\varphi(\boldsymbol{x}) +\theta_z  \,. 
	\label{eq:CenterTrafosAlg} 
\end{align}
Moreover, gauge transformations can be used to rotate $\varphi(\boldsymbol{x})$ into the Cartan subalgebra and we arrive at 
\begin{align}
	\nu(\boldsymbol{x}) = \varphi_3(\boldsymbol{x})\, t^3+ \varphi_8(\boldsymbol{x}) \,t^8\,, 
\end{align}
with the eigenvalue fields $\nu_i(\boldsymbol{x})$, 
\begin{align}
	\nu_{\pm}= \frac12 \left( \varphi_8 \pm \frac{1}{\sqrt{3}}\, \varphi_3\right)\,,\qquad \nu_3= -\varphi_8 \,, 
	\label{eq:nuField}
\end{align}	
and, with a slight abuse of notation 
\begin{align} 
	\varphi^c = \nu_+ P_+ + \nu_- P_- + \nu_3 P_3\,,
	\label{eq:varphicinvau}
\end{align}
where the matrices $P_A$ are simply projection matrices $P_A^2 =P_A$ on the $A$'s entry of the diagonal, 
\begin{align} 
 (P_A)^{BC}=\delta^{AC}\delta^{AB}\,.
 \label{eq:PA}
\end{align}
Note that the $\nu_A$ over-determine the $\varphi_i$, as can be seen from \labelcref{eq:nuField}. This will be used to our advantage in \Cref{app:DSE-Polya}, where the Polyakov loop potential is computed in terms of potentials for the eigenvalues in the adjoint representation. 

With \labelcref{eq:nuField}, the traced Polyakov loop \labelcref{eq:TracedPLoop} takes the form 
\begin{subequations} 
		\label{eq:PbarPLoopAlg}
\begin{align}
	{L}(\varphi_3,{\varphi}_8) = \frac{1}{3}\left[e^{- \frac{2 \pi}{\sqrt{3}}\imag {\varphi}_8} + 2 \, e^{\frac{\pi}{\sqrt{3}} \imag {\varphi}_8} \cos{\pi\varphi_3}\right]\,, 
	\label{eq:PLoopAlg}
\end{align}
while the traced loop $\bar L$ in \labelcref{eq:TracedPLoop} reads 
\begin{align}
	\bar{L}(\varphi_3,{\varphi}_8) = \frac{1}{3}\left[e^{ \frac{2 \pi}{\sqrt{3}} \imag{\varphi}_8} + 2 \, e^{-\frac{\pi}{\sqrt{3}} \imag {\varphi}_8} \cos{\pi\varphi_3}\right]\,. 
	\label{eq:barPLoopAlg}
\end{align}
\end{subequations} 
\Cref{eq:PLoopAlg} can be rewritten in terms of the $\nu_i(\boldsymbol{x})$, using e.g.\ 
\begin{align}
  \varphi_3= \sqrt{3}\left( \nu_+ - \nu_-\right) \,,\qquad 	\varphi_8=  - \nu_3 \,. 
	\label{eq:varphi38nui}	
\end{align}
\Cref{eq:varphi38nui} make the underlying gauge invariant nature of \labelcref{eq:PLoopAlg} apparent. The center shifts of $\nu_i$ following from that of the algebra fields \labelcref{eq:CenterTrafosAlg} as  
\begin{subequations}
	\label{eq:CenterEigenvalues}
	\begin{align}
		\nu_i \to \nu_i+\omega_i\,,
		\label{eq:Centernu}
	\end{align}
	with 
	\begin{align}
		\omega_\pm = 0\,,\, \frac{2}{\sqrt{3}} \,,  -\frac{1}{\sqrt{3}}\,,\qquad  \omega_3 = 0\,,\,1\,.
		\label{eq:Centeromegai}
	\end{align}
\end{subequations}
The fixed points of the combination of a center transformation \labelcref{eq:Centeromegai} with a Weyl reflection are the zeros of ${L}(\varphi_3,{\varphi}_8)$, see e.g.~\cite{Herbst:2015ona}. 

In summary we conclude that their expectation values  $\nu_i = \langle \hat \nu_i\rangle $ serve as order parameters for center-symmetry breaking. Here, the $\hat{\ }$ indicates field operators and we reserve $\nu,\varphi,...$ for the mean fields.  The respective symmetry breaking information can conveniently be combined in a single order parameter, the traced Polyakov loop of $\langle \nu(\boldsymbol{x})\rangle$. 
It can be written as  
\begin{subequations} 
	\label{eq:Lnu} 
	\begin{align}
		L(\langle \varphi_3\rangle\,,\, \langle \varphi_8\rangle ) := 	L(\langle \nu\rangle )\,,
		\label{eq:Lnu1}
	\end{align}
	with 
	\begin{align}
	 \langle \varphi_3\rangle = \sqrt{3}\left(  \langle \nu_+\rangle -\langle \nu_-\rangle\right) \,,\qquad 	\langle \varphi_8\rangle =  -\langle \nu_3\rangle \,, 
		\label{eq:Lnu2}	
	\end{align}
\end{subequations} 
using \labelcref{eq:varphi38nui}. It is left to compute the gauge invariant expectation values $\nu_i$ within the DSE approach used in the present work. This is detailed in \Cref{sec:Conf-Deconf} and \Cref{app:DSE-Polya}. 

We close this discussion with the remark, that it has been shown in \cite{Herbst:2015ona} that the difference between $\langle L\rangle $ and $L(\langle \nu\rangle )$ is a thermal normalisation factor 
that diffuses the symmetry breaking information of the order parameter already in pure Yang-Mills theory: while $L(\langle \nu\rangle )$ rapidly converges to unity in the deconfined regime with $T\gtrsim 1.3\, T_\textrm{conf}$, the order parameter $\langle L\rangle $ converges very slowly and even overshoots unity for a very large regime. Finally, while the connection of center-symmetry breaking and the confinement-deconfinement phase transition is a direct one in pure Yang-Mills, this connection is less obvious in QCD, where center-symmetry breaking is described by a soft crossover and the definition of the confinement-deconfinement transition is still under debate. We refrain from entering this debate here and simply use the center symmetry crossover as a proxy for the confinement-deconfinement phase transition.

%%%%%%%%%%%%%%%%%%%%%%%%%%%%%%%%
\subsection{Eigenvalue potential}
\label{app:DSE-Polya}

In \Cref{app:PloopAlg} we have reviewed the functional approach to center-symmetry breaking. This is based on the gauge invariant eigenvalue fields of the Polyakov loop or rather the traced Polyakov loop of the expectation values of the eigenvalue fields as an order parameter, see \labelcref{eq:PbarPLoopAlg,eq:Lnu}.  It is left to compute these expectation values within the DSE approach. We follow the background field approach to the Polyakov loop potential put forward in \cite{Fister:2013bh} for general functional approaches. For more details we refer to this work and references therein.

%%%%%%%%%%%%%%%%%%%%%%%%%
\subsubsection{Background field approach} 
\label{app:BackgroundEffAction}

In the background field approach the gauge field is split into an auxiliary background field $\bar A$ and a dynamical fluctuation field $a_\mu$. Then, the partial derivative in the covariant gauge is turned into the covariant one with the background field, 
\begin{align} 
	\partial_\mu A_\mu =0 \to \bar D_\mu a_\mu=0\,,\qquad A_\mu=\bar A_\mu+a_\mu\,, 
	\label{eq:Backgroundgauge} 
\end{align}
with the covariant derivative 
\begin{align}
	{D}_{\mu} = \partial_{\mu} - \imag g_s {A}_{\mu}\,, 
	\label{eq:CovDer}
\end{align}
and $\bar D=D(\bar A)$.  The respective classical gauge fixed action $S[\bar A,\Phi]$ underlying the DSEs is given by, 
\begin{subequations} 
		\label{eq:ClassicalAction}
\begin{align}\nonumber 
	S_\textrm{QCD}[\bar{A},\phi] =& \int_x  \left[ \bar{q}(\Drbar- \gamma_0\mu_q +m_q)q + \frac{1}{4} F_{\mu\nu}^{a} F_{\mu\nu}^{a} \right] \\[2ex]
	 & \,+ S_\textrm{gf}[\bar A,a]+ S_\textrm{gh}[\bar A,a,c,\bar c]\,, 
	 	\label{eq:ClassicalActionPhys}
\end{align} 
where $\Phi=(a_\mu, c,\bar c, q, \bar q)$ is the fluctuation superfield, see \labelcref{eq:SuperField}. The current quark mass matrix and the quark chemical potential are diagonal 
\begin{align} 
	m_q=\textrm{diag}({m_u\,,\,m_d\,,\,m_s)}\,,\qquad \mu_q =\textrm{diag}({\mu_u\,,\,\mu_d\,,\,\mu_s)}\,,
\label{eq:CurrentMass+muq}
\end{align} 
In the present work we consider $\mu_u=\mu_d=\mu_s=\mu_B/3$,  see 	\labelcref{eq:muqs}. Moreover, the computations are performed in the isospin-symmetric approximation, $m_u=m_d=m_l$. 

The second line in 	\labelcref{eq:ClassicalActionPhys} constitutes the gauge fixing sector with 
\begin{align}	 \nonumber 
	S_\textrm{gf}[\bar A,a] =&\, \frac{1}{2\xi} \int_x \left( \bar{D}_{\mu} a_{\mu} \right)^2 \,,\\[1ex] 
	S_\textrm{gh}[\bar A,a,c,\bar c] =&- \int_x  \bar{c}^a \bar{D}_{\mu} D_{\mu}^{ab} c^{b} \,,
	\label{eq:ClassicalActionGauge}
\end{align}
\end{subequations} 
The covariant derivative \labelcref{eq:CovDer} in the adjoint representation has the form 
\begin{align}
	D_\mu^{bc} = \delta^{bc}\partial_\mu - g_s A^a_\mu f^{abc}\,, 
	\label{eq:CovDerAd}
\end{align}
with $(t^a)^{bc} = -i \, f^{abc}$. The respective effective action depends on the background field and the fluctuation field separately, 
\begin{align}
\Gamma= \Gamma[\bar A ,\phi]\,\qquad \phi=(a,c,\bar c,q,\bar q)\,,
\end{align}
where $\phi$ is the superfield of all dynamical fields, including not only the fluctuation gauge field $a$, but also the ghost and anti-ghost $c,\bar c$ and the quarks and anti-quarks $q,\bar q$. In the present isospin symmetric 2+1 flavour QCD setup we have $q=(l,s)$ with the light quark field $l=(u,d)$. 

The construction admits the definition of a gauge invariant effective action 
\begin{align}
	\Gamma[A] = \Gamma[A,0]\,, 
	\label{eq:GaugeInvGamma} 
\end{align} 
A priori, the gauge invariance of $\Gamma[A] $ is that of the auxiliary background field. However, it indeed encodes the physical one as the respective Ward identities are related to the STIs of the full gauge field via the Nielsen identities. Part of \labelcref{eq:GaugeInvGamma} is the gauge invariant Polyakov loop potential, expressed in terms of $\varphi$, 
\begin{align}
	\Gamma[A] = \int_x V_\varphi(\varphi) + \cdots \,, 
\end{align}
where in an abuse of notation we use $\varphi$ also for its mean field. Evidently, due to its gauge invariance, the Polyakov loop potential $V_\varphi$ can be expressed in terms of $\langle \nu\rangle$ or the respective mean fields $\varphi_{3,8}$ in \labelcref{eq:Lnu2} by simply rotating $\varphi$ into the Cartan subalgebra. For constant fields this implies \labelcref{eq:nuA} with the components 
\begin{align}
\varphi_{3,8} = \frac{g_s \beta}{2 \pi} A_0^{3,8}\,. 
\label{eq:varphiA0}
\end{align}
for the dimensionless fields $\varphi_{3,8}$. The respective dimensionless effective potential is given by 
\begin{align}
	 V_\varphi(\varphi^c) = \frac{1}{{ T^3 \cal V}_3}\Gamma[A_0(\varphi),\boldsymbol{A}=0]\,,
\label{eq:DefofVvarphi} 
\end{align}
with the spatial volume ${\cal V}_3$. In \labelcref{eq:DefofVvarphi} we divided out the four-dimensional volume $\beta {\cal V}_3$ as well as the overall explicit dimensional factor $T^4$. 
A given mean field can be seen as the solution of the EoM with an external current, 
\begin{align} 
	 \frac{\partial V_\varphi}{\partial \varphi^c} = J_\varphi^c\,,  
	\label{eq:EoMJ0}
\end{align}
with the Cartan components of the constant $J_\varphi$-current, given by $2 \pi/(g_s \beta) J^{3,8}_0$ and $J_0$ is the zero component of the gauge field current. The physical expectation value is obtained for $J_\varphi=0$.

%%%%%%%%%%%%%%%%%%%%%%%%%
\subsubsection{Background DSE for the Eigenvalue potential} 
\label{app:DSEA0}

\begin{figure}[t]
	\centering
	\includegraphics[width=0.96\columnwidth]{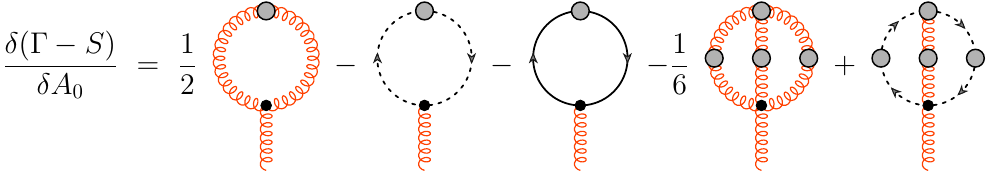} 
	\caption{Functional background field DSE. Full propagators and vertices are indicated by grey blobs, the classical vertices are indicated by small black blobs. Gluons are represented by red spiral lines, ghosts by back dotted ones, and quark by straight black ones. In contradistinction to the fluctuation field DSE, the background field DSE also hosts a two-loop ghost-gluon term with the four-point vertex $S^{(4)}_{ac\bar c \bar A_0}$. }
	\label{fig:DSE-A}
\end{figure}
\Cref{eq:DefofVvarphi} allows us to determine the effective potential $V_\varphi$ or rather its derivative from the respective functional background field DSE \cite{Fister:2013bh}. Schematically the latter reads 
\begin{align}\nonumber 
	\frac{\delta \left( \Gamma - S_A \right) }{\delta A_0}
	=& \, \frac{1}{2} S_{aaA_0 }^{(3)} G_{aa} - S_{A_0 c\bar{c}}^{(3)} G_{c \bar c} - S^{(3)}_{q\bar{q}A_0 } G_{q \bar q} \\[1ex]
	& \hspace{-1cm}- \frac{1}{6} S_{
		aaaA_0 }^{(4)} G_{aa}^3 \Gamma_{aaa}^{(3)} + S^{(3)}_{a c\bar{c}A_0 } G_{c \bar c }^2 G_{aa}\Gamma_{ac\bar{c}}^{(3)}\,,  
	\label{eq:A0DSE}
\end{align}
with the gluon, ghost and quark propagators $G_{aa}, G_{c\bar c}, G_{q,\bar q}$. Moreover, in \labelcref{eq:A0DSE} we have used the notation 
\begin{align} 
F^{(n+m)}_{\phi_{i_1}\cdots \phi_{i_n} \bar A^m}[\bar A, \phi] = \frac{F[\bar A,\phi]}{\delta \phi_{i_1} \dots \phi_{i_n}\delta \bar A^m}\,, 
\label{eq:FunDerivatives}
\end{align}
for functional derivatives with respects to the fields $\phi$ and $\bar A_\mu$. In \labelcref{eq:A0DSE}, the propagators are contracted in momentum, Lorentz and internal indices with the vertices, a pictorial description of \labelcref{eq:A0DSE} is given in \Cref{fig:DSE-A}. 

\begin{figure}[t]
	\centering
	\includegraphics[width=0.85\columnwidth]{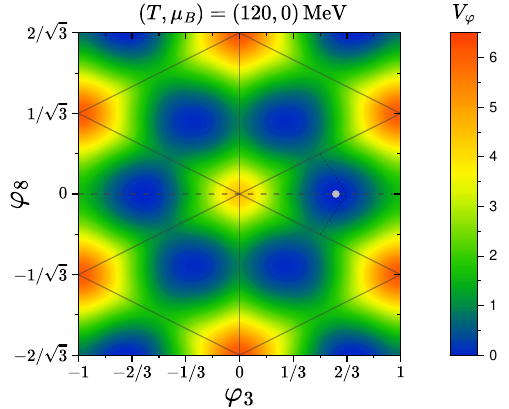}
	\caption{Heat map of the Polyakov loop potential $V_{\varphi}(\varphi_3,\varphi_8)$, see \labelcref{eq:DefofVvarphi}, in the $\varphi_3$ - $\varphi_8$ plane. We have normalised the value of the potential at the minimum to zero.}
	\label{fig:V38_contour_120_0}
\end{figure}

Yang-Mills theory is center-symmetric and the potential is also invariant under Weyl reflections, see e.g.~\cite{Herbst:2015ona}. 
Then, $\varphi_8$ can be set to zero and \labelcref{eq:PLoopAlg} takes real values. This choice picks up one of the minima of the potential, and a similar choice is done on the lattice, see the contour plot \Cref{fig:V38_contour_120_0}. 
In QCD with vanishing chemical potential, the choice $\varphi_8=0$ is still possible, keeping a real Polyakov loop $L(\langle \nu\rangle)$ in \labelcref{eq:Lnu}, see e.g.~\cite{Braun:2009gm, Fischer:2013eca, Fischer:2014vxa, Fu:2015naa, Reinosa:2015oua}. 
In both situations we have $L(\langle \nu\rangle)= \bar L(\langle \nu\rangle)$, reflecting $\langle L\rangle^\dagger =  \langle L^\dagger\rangle$ at $\mu_B=0$.  

\begin{figure}[t]
	\centering
	\includegraphics[width=0.85\columnwidth]{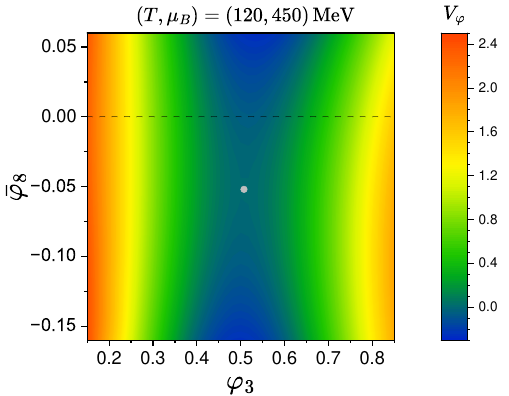} 
	5
	\caption{Heat map of the the Polyakov loop potential $V_{\varphi}(\varphi_3,\varphi_8)$, see \labelcref{eq:DefofVvarphi}, in the $\varphi_3$ - $\bar \varphi_8$ plane, with $\varphi_8= \imag \bar\varphi_8$. We have normalised the value of the potential at the minimum to zero.}
	\label{fig:V38_contour_120_450}
\end{figure}

At finite chemical potential this symmetry is broken and we have $\langle L\rangle^\dagger \neq   \langle L^\dagger\rangle$ due to the medium. Indeed, for positive $\mu_B$ we find 
\begin{align}
	|\langle L\rangle| \leq 	|\langle L^\dagger\rangle|\,. 
	\label{eq:ExpLLdagger}
\end{align} 
Moreover, for $\mu_B\neq 0$, the choice $\varphi_8=0$ does not constitute a minimum or saddle point of the Polyakov loop potential any more, see the contour plot  \Cref{fig:V38_contour_120_450}. 
Instead, we find saddle points with $\varphi_3 \in \mathbbm{R}$ and $\varphi_8\in \imag \,\mathbbm{R}$, that also lead to a real Polyakov loop $L(\langle \nu\rangle)$, see \labelcref{eq:PLoopAlg}. 
For a recent comprehensive discussion see \cite{Maelger:2017amh,Reinosa:2015oua}. These properties reflect the finding in low energy effective theories with the Polyakov loop, for respective discussions see e.g.~\cite{Dumitru:2005ng, Schaefer:2007pw} and in particular the review \cite{Fukushima:2017csk}. 
Specifically, it has been shown in \cite{Schaefer:2007pw}, that the construction of model potentials that still allow for minima for the equations of motion for $\langle L\rangle $, and hence for $\langle \nu\rangle$, is incompatible with physics constraints.  

The mean value $\langle \nu\rangle$ is derived from the solution of the EoM for $\varphi_{3,8}$ in 	\labelcref{eq:EoMJ0} for vanishing external current $J_\varphi=0$, 
\begin{align}
	\left. \frac{\partial V_\varphi(\varphi_3,\varphi_8)}{\partial \varphi_i}\right|_{\varphi_i=\langle \varphi_i\rangle} =0\,,\qquad i=3,8\,. 
	\label{eq:EoMVvarphi}
\end{align}
The solutions of the EoMs are the expectation values of the algebra fields, $\varphi^\textrm{EoM}_i =\langle \varphi_i\rangle$. The expectation value $\langle \nu\rangle$ follows with \labelcref{eq:nuField,eq:Lnu2}. It is left to compute $V_\varphi(\varphi_3,\varphi_8)$, and for this purpose we resort to a modification of the optimised approximation scheme discussed in \cite{Fister:2013bh}. There, the renormalisation point in Yang-Mills theory was chosen such that the impact of the two-loop terms is minimised around the phase transition. In the present work we choose the renormalisation point $\mu_\textrm{\tiny{RG}} = 40$\,GeV of 2+1 vacuum QCD computation in \cite{Gao:2021wun}, underlying the computation here. Then it is checked that this indeed minimises the contributions of the two-loop terms by the benchmark computations of fluctuations of conserved charges at $\mu_B=0$: our results agree well with the respective lattice results, see \Cref{fig:chiB-zeromu}. 

We proceed with the explicit computation of the Polyakov loop potential in this approximation. The first two terms on the right hand side of \labelcref{eq:A0DSE}, the gluon and ghost loop, constitute 
the $A_0$-derivative of the pure glue potential, $V_{\textrm{gl}}$, while the third term is the $A_0$-derivative of the quark part of the potential,  $V_q$, to wit, 
\begin{align}
	V_\varphi(\varphi_3,\varphi_8) = V_{\textrm{gl}}(\varphi_3,\varphi_8) + V_q(\varphi_3,\varphi_8)\,,
	\label{eq:potential-1L}
\end{align}
with the split of the glue potential in the gluon and ghost loop parts 
\begin{align}
	V_{\textrm{gl}}(\varphi_3,\varphi_8) = V_a(\varphi_3,\varphi_8)+V_c(\varphi_3,\varphi_8)\,.
	\label{eq:Vgl-Va+Vc}
\end{align}
Accordingly, the numerical computation of the potential is done in terms of the ghost, gluon and quark potentials and we illustrate this computation in detail at the example of the gluon potential. 

%

%%%%%%%%%%%%%%%%%%%
\subsubsection{Computation of gluonic potential $V_a(\varphi_3,\varphi_8)$} 
\label{app:Va}

Here we discuss the computation of the different parts of the full potentials at the example of $V_a$: to begin with, the gluon loop contains a colour trace in the adjoint representation and the respective eigenvalues are given by 
\begin{subequations}
	\label{eq:Eigenvalues+vectorsAd}
\begin{align}
	\varphi^{(\textrm{ad})}(\boldsymbol{x}) \bigl|\psi^{(\textrm{ad})}_{\nu_i}\bigr\rangle = \nu^{(\textrm{ad})}_i(\boldsymbol{x}) \bigl|\psi^{(\textrm{ad})}_{\nu_i}\bigr\rangle \,, \qquad
	\label{eq:Eigenvalues+vectorsAd1}
\end{align}
with $i=1,...,8$ and the eigenvalues in the adjoint representation 
\begin{align}
	\nu^{(\textrm{ad})}_i\in \left( 0\,,\, 0\,,\, \pm \varphi_3\,,\, \pm \frac{\varphi_3\pm\sqrt{3}\,\varphi_8}{2}\right)\,.  
\label{eq:Eigenvalues+vectorsAd2}
\end{align}
\end{subequations} 
For the computation of the effective potential we consider constant $\varphi_{3,8}$ and hence constant eigenvalues $\nu_A$ and $\mu_a$. For the sake of convenience we use the representation of the Cartan field $\varphi^c$ in the adjoint representation, 
\begin{align} 
	\varphi^{c,\textrm{(ad)}} = \sum_{i=1}^8 \nu^{(\textrm{ad})}_i P_i\,, \qquad P_i^2 =P_i\,,  
	\label{eq:varphicnuad}
\end{align} 
the analogue of \labelcref{eq:varphicinvau,eq:PA} in the adjoint representation. We have dropped the contributions of the first to eigenvalues as they are vanishing, see \labelcref{eq:Eigenvalues+vectorsAd2}. As in the fundamental representation, 	\labelcref{eq:varphicnuad} is over-complete. However, we shall use this over-completeness to our advantage: we can turn the $\varphi_{i}$-derivatives of $V_a(\varphi_3,\varphi_8)$ in that with respect to the eigenvalue field $\nu^{(\textrm{ad})}_i$,  
\begin{align} 
	S_{aaA_0 }^{(3)}  \to \frac{\partial S^{(2)}_{aa} }{\partial {\nu^{(\textrm{ad})}_j}}\,.
	\label{eq:EigenvalueDerivatives} 
\end{align} 
The derivatives with respect to $\varphi_{3,8}$ and hence that with respect to $A_0$ follow as linear combinations of the eigenvalue derivatives, using \labelcref{eq:varphicnuad}. The benefit of \labelcref{eq:EigenvalueDerivatives} is the fact, that the right hand side of \labelcref{eq:EigenvalueDerivatives}  lies in the eigenspace of $\nu^{(\textrm{ad})}_j$. This facilitates the computation of the colour trace in the gluon loop in \labelcref{eq:A0DSE}. 

The loop also involves the gluon propagator, which is also diagonal: to begin with, it admits an orthogonal split in chromo-electric, chromo-magnetic and gauge part, 
\begin{align} \nonumber 
	G_{aa}(p) = &\,\frac{1}{Z_A^E(x) x\, } \Pi^{E}(p) + \frac{1}{Z_A^M(x) x\, } \Pi^{M}(p)\\[1ex] 
	&\, + \frac{\xi}{x}\, \Pi^L(p)\,. 
	\label{eq:Gaa}
\end{align}
with the covariant momentum $p=p(\varphi^c)$ and the covariant Laplacian $x=x(\varphi^c)$, 
\begin{align} 
	p(\varphi^c) =&\, (\omega_n + 2 \pi \varphi^c\,,\,\boldsymbol{p})\,, \qquad x(\varphi^c)= p_\mu^2\,.
	\label{eq:p2}
\end{align}
In 	\labelcref{eq:Gaa} we have used constant backgrounds $\varphi^c$ and in such a background the projection operators $ \Pi^{E,M},\Pi^L$ take the form 
\begin{align} \nonumber 
	\Pi^{E}(p) = &\,(1-\delta_{\mu0})(1-\delta_{\nu0})\left(\delta_{\mu\nu} - \frac{p_{\mu} p_{\nu}}{\boldsymbol{p}^2}\right)\,,  \\[1ex]
	\Pi^{M}(p) = &\, \Pi^{\bot}(p) - \Pi^{E}(p)\,,   
		\label{eq:ProjectionsT}
\end{align}
with the transverse and longitudinal vacuum projection operators 
\begin{align} 
	\Pi^\bot(p) = \,\delta_{\mu\nu} - \frac{p_\mu p_\nu}{x(\varphi^c)} \,, \qquad    \Pi^L =\mathbbm{1}- \Pi^\bot \,.
	\label{eq:Projections}
\end{align}
Finally we show that the gluon propagator is diagonal in the coordinate system, spanned by the eigenvectors of $\varphi^c$. We also perform the trace over Lorentz indices with $G_{aa}^{\mu\nu}\delta^{\mu\nu} \to G_{aa}^{\mu\mu}$ and arrive at 
\begin{align} \nonumber 
\left\langle\psi^{(\textrm{ad})}_{\nu_i}\right| G^{\mu\mu}_{aa}(p)  \left|\psi^{(\textrm{ad})}_{\nu_j}\right\rangle &= \delta_{ij}  \\[1ex]
&\hspace{-3.3cm}\times \Biggl[\frac{1}{Z_A^E(x^{(\nu_i)}) x^{(\nu_i)}} + \frac{2}{Z_A^M(x^{(\nu_i)}) x^{(\nu_i)}\, }+ \frac{\xi}{x^{(\nu_i)}}\Biggr]\,,
\label{eq:DiagonalGaa}
\end{align} 
where we have used $\Pi^E_{\mu\mu}=1$, $\Pi^M_{\mu\mu}=2$ and $P^L_{\mu\mu}=1$. 
In \labelcref{eq:DiagonalGaa} we have used the covariant momentum $p^{(\nu_i)}$, which carries no colour structure, 
\begin{align}
p  \left|\psi^{(\textrm{ad})}_{\nu_j}\right\rangle  =p^{(\nu_i)} \left|\psi^{(\textrm{ad})}_{\nu_j}\right\rangle\,,
\label{eq:p2nu}
\end{align}
with 
\begin{align}
 p^{(\nu_i)} = (\omega_n + 2 \pi \nu^{(\textrm{ad})}_i\,,\,\boldsymbol{p}) \,,
	\label{eq:p2nu1}
\end{align}
and the respective covariant Laplacian $x^{(\nu_i)}$, 
\begin{align} 
 	x^{(\nu_i)}  = (2 \pi T)^2 (n+\nu_i^{(\textrm{ad})})^2 + \boldsymbol{p}^2\,.  
 	\label{eq:Laplacenu}
\end{align}
Note that \labelcref{eq:p2nu1,eq:Laplacenu} are colour scalars and do not carry a colour structure such as $x(\varphi^c)$ in \labelcref{eq:p2}. With these preparation it also follows, that  \labelcref{eq:EigenvalueDerivatives} reads 
\begin{align}
	\frac{\partial S^{(2)}_{aa} }{\partial {\nu^{(\textrm{ad})}_j}} = (2 \pi T)^2(n+\nu_i^{(\textrm{ad})}) \,P_{i} \,,
	\label{eq:EigenvalueDerivativesExplicit} 
\end{align} 
with the projection operator $P_{i}$  on the eigenspace of $\nu_i^{(\textrm{ad})}$. Putting everything together, we are led to 
\begin{widetext} 
\begin{align}
	\frac{\partial V_\textrm{gl} }{\partial {\nu^{(\textrm{ad})}_j}}= \frac12\frac{1}{T^3 {\cal V}_3} \,\Tr \, \frac{\partial S^{(2)}_{aa} }{\partial {\nu^{(\textrm{ad})}_j}}  G_{aa} =  \sum_{i=1}^8\left[	\frac{1}{2T} \sum_{n\in\mathbbm{Z} }\int \!\! \frac{d^3 p}{(2 \pi)}(n+\nu_i)   \left\{\frac{1}{Z_E(x^{(\nu_i)}   ) \,x^{(\nu_i)}  } + \frac{2}{Z_M(x^{(\nu_i)}  )\, x^{(\nu_i)}  }+\frac{1}{ x^{(\nu_i)} } \right\}\right]\,.
	\label{eq:Vmode1} 
\end{align}
\end{widetext} 
With \labelcref{eq:Vmode1} the computation of the full glue part of the potential can be separated into the computation of three mode potentials $V_{\textrm{mode}}(\varphi)$ for the chromo-electric, chromo-magnetic and gauge modes for a scalar shift $\varphi$ of the Matsubara frequency. These mode potentials or rather their $\varphi$-derivatives can be defined for all fields,  
\begin{subequations}
	\label{eq:VmodeComp}
\begin{align} 
	\frac{\partial V_{\textrm{mode},\phi_i}(\varphi)}{\partial \varphi} = \frac{s_{\phi_i}}{2}\frac{1}{T}\sum_{n\in \mathbbm{Z}}\int \frac{d^3 p}{(2\pi)}  \, (n+ \varphi) f_{\phi_i}(x^{(\varphi)})\,, 
	\label{eq:VmodeIntegral}
\end{align}
with $x^{(\varphi)}$ in \labelcref{eq:Laplacenu} and 
\begin{align}
	\left( s_{a^E}\,,\, s_{a^M}\,,\,s_{a^L} \,,\,s_{c} \,,\,s_{q} \right) =(1,2,1,-2,-2)\, .
\end{align}
\end{subequations}
The prefactor $s_{\phi_i}$ takes care of the relative minus sign of the fermionic loops as well as the multiplicity of all field modes. All internal traces have been already performed in  \labelcref{eq:VmodeIntegral}, which is a simple numerical sum and integral. The integrands $f_{\phi_i}$ of the fields $\phi=(a^\bot\,,\,a^{L}\,,\, c\,,\,q)$ are  $f_a^{\bot}\,,\,f_a^{L}\,,\,f_c\,,\,f_q$ and are provided the scalar parts of the gluon, ghost and quark propagators computed from the respective DSEs. They are provided in  \labelcref{eq:ModeIntegrandsfa} (gluons), \labelcref{eq:ModeIntegrandsfc} (ghosts), and \labelcref{eq:ModeIntegrandsfq} (quarks).  For the chromo-electric, chromo-magnetic and longitudinal gluons they are given by 
\begin{align}
	f_{a^{E,M}}(x) = \frac{1}{Z_A^{E,M}(x)}\,,\qquad  f_{a^{L}}(x) = \frac{1}{x}\,.
	\label{eq:ModeIntegrandsfa} 
\end{align}
Finally, the full gluon part of the Polyakov loop potential is obtained from the mode potentials with the sum over all modes and all eigenvalues 
\begin{align}
	V_a(\varphi_3,\varphi_8) = \sum_{A=E,M,L} V_{a^A}(\varphi_3,\varphi_8)\,, 
\label{eq:VPol-EigenvalueSum}
\end{align} 
with 
\begin{align}
V_{a^A}(\varphi_3,\varphi_8)=&\,	3 \sum_{i=1}^8  V_{\textrm{mode},{a^A}}(\nu_i)\,,
\end{align}
with $A=E,M,L$.  The longitudinal part can be computed analytically, it is simply half of the full one-loop potential in Yang-Mills theory, \cite{Gross:1980br, Weiss:1980rj}. With $V_{a^L}=V_\textrm{\tiny{GPY-W}}/2$ it follows 
\begin{align} 
V_{\textrm{mode},a^L}(\varphi)  \simeq 
	\frac{\pi^2}{24} \left[ 4 \left( \tilde \varphi -\frac12\right)^2 -
	1\right]^2\hspace{-.2cm}\,, \quad \tilde \varphi ={ \varphi\ \textrm{mod}\, 1}\,. 
\end{align} 
This concludes the step-by-step evaluation of the computation of the gluonic part of the Polyakov loop potential.

%%%%%%%%%%%%%%%%%%%%%%%%%
\subsubsection{Computation of the ghost potential $V_c(\varphi_3,\varphi_8)$} 
\label{app:Vc}

All steps are generic and the computation readily extends to the ghost part of the glue potential. The ghost integrand in the mode potential relation 	\labelcref{eq:VmodeIntegral} is given by  
\begin{align}
f_c(x)=  \frac{ 1}{Z_c(x) x}\,, \qquad \Gamma_{c\bar c}^{(2)} = Z_c(x)\, x\, \mathbbm{1}_\textrm{ad}\,, 
\label{eq:ModeIntegrandsfc}
\end{align}
with $\mathbbm{1}_\textrm{ad}$ in the adjoint representation.

%%%%%%%%%%%%%%%%%%%%%%%%%
\subsubsection{Computation of the quark potential $V_q(\varphi_3,\varphi_8)$} 
\label{app:Vq}

The quark part $V_q$ of the potential follows similarly. However, the quark carries the fundamental representation and hence we have 
\begin{align} 
	V_q(\varphi_3,\varphi_8) =\sum_{\alpha=\pm,3} V_{\textrm{mode},q}(\nu_\alpha)  \,,
	\label{eq:Vq-EigenvalueSum}
\end{align} 
with the eigenvalues in the fundamental representation provided in \labelcref{eq:nuField}. The integrand $f_q$ of quark mode potential \labelcref{eq:VmodeIntegral} is given by \labelcref{eq:ModeIntegrandsfqMain}, and we recall it here for the sake of completeness, 
\begin{align}
	f_q(p_{q,\varphi},\mu_q)=  \frac{1}{Z_q ^E} \frac{1}{ p_{q,0}^2+\left(\frac{Z_q^M}{Z_q^E}\right)^2\boldsymbol{p}^2+M^2_q} \,.
	\label{eq:ModeIntegrandsfq}
\end{align}
The matrix valued quark momentum \labelcref{eq:pqMain} reduces to 
\begin{align} 
		p_{q} \to  (\omega_{q,n} + 2 \pi \varphi- \imag \mu_q \,,\,\boldsymbol{p})\,,
	\label{eq:Eigenvaluepq}
\end{align} 
with the eigenvalue $\varphi$ instead of the colour matrix $\varphi^c$ in \labelcref{eq:pqMain}. Evidently, 
the quark momenta $p_q$ are different from that of the gluons and the ghosts in \labelcref{eq:p2,eq:Laplacenu}: 
we have to accommodate the quark chemical potential $\mu_q$ as well as the quark Matsubara frequencies $\omega^q_n = 2 \pi T(n+1/2)$. 
Furthermore, all quark dressings $Z=Z_q^{E,M},M_q$ depend on $p_q$ and $\mu_q$ separately, 
\begin{align} 
	Z=Z(p_q,\mu_q)\,.
	\label{eq:Zqxqmuq}
\end{align}
This parameterisation is amiable to the Silver blaze property: at $T=0$ and below the baryonic onset chemical potential, $\mu_q< \mu_q^\textrm{(on)}$, the dressings only depend on $x_q$: their imaginary part originates solely in the imaginary part of the complex variable $x_q$. 
In turn, for $\mu_q> \mu_q^\textrm{(on)}$, the dressings develop a genuine $\mu_q$-dependence. 

In the present work we consider temperatures $T\gtrsim 100$\,MeV. 
Instead of \labelcref{eq:Zqxqmuq} we use the approximation 
\begin{align} \nonumber 
	Z\approx &\,  \textrm{Re}(Z)(p_q(\mu_q=0)\,,\,\mu_q)\\[1ex] 
	&\hspace{1cm}+ \imag\, \textrm{Im}(Z)(p_q(\mu_q=0),\mu_q)\,.
	\label{eq:Zqapprox}
\end{align} 
The split of the dressings $Z_q^{E,M},M_q$ in their imaginary and real parts suits the numerics and suffices for the present purpose. This concludes the discussion of the numerical computation of the Polyakov loop potential.

%%%%%%%%%%%%%%%%%%%%%%%%%
\subsubsection{Eigenvalue potential and Eigenvalues at finite $\mu_B$} 
\label{app:Vvarphi-muB}

In summary, the glue part of the Polyakov loop potential  in the present approximation still carries the center symmetry while the quark breaks it explicitly. Consequently, the center-symmetry phase transition in Yang-Mills theory turns into a rather soft crossover in full QCD with dynamical quarks. Moreover, at finite chemical potential, the quark part of the potential is complex for non-vanishing $\varphi_8$: $V_q(\varphi_3,\varphi_8\neq 0)\in \mathbbm{C}$, while it is real for $\varphi_8=0$. Hence, the respective effective potential and the effective action are functionals of the complex variable $\varphi_8$ as discussed in detail in \cite{Ihssen:2022xjv}: the Legendre transform with respect to $A_0$ is defined at vanishing $\mu_q$ and the effective action at $\mu_B\neq 0$ is the analytic continuation of that at $\mu_q=0$. The latter is defined by the functional relation for $\Gamma[\bar A,\phi]$ itself, either the DSE as here or the fRG as discussed in \cite{Ihssen:2022xjv}. This entails, that the effective potential depends on the complex fields $\varphi_i$, but not their complex conjugate $ \varphi_i^*$. It satisfies the Cauchy-Riemann equations 
\begin{align} 
	\frac{\partial V_\varphi(\varphi_3,\varphi_8)}{\partial \varphi^*_i}=0\,, \qquad i=3,8\,,
	\label{eq:CR-V_varphi}
\end{align}
and is computed as detailed above also for $\mu_q\neq 0$. \Cref{eq:CR-V_varphi} has to be paired with \labelcref{eq:EoMVvarphi} for a computation of the solutions $\varphi_{3,8}^\textrm{EoM}$ of the EoMs at finite chemical potential. Moreover, the solution of the equations of motion still define the vanishing  current $J_\phi=0$ required for the definition of correlation functions. At finite $\mu_q$ the solutions to the EoMs are now saddle points and not minima, see 	\Cref{fig:V38_contour_120_450}. Moreover, taking the $\mu_q$-derivative of the EoMs \labelcref{eq:EoMVvarphi}, one can show that 
\begin{align} 
\partial_{\mu_q} \langle \hat \varphi_3\rangle \in \mathbbm{R}\,,\qquad  \partial_{\mu_q} \langle \hat \varphi_8\rangle \in \imag\, \mathbbm{R}\,. 
\end{align} 
More explicitly, the $\mu_q$-derivative of the EoM leads to the relations  
\begin{align}\nonumber  
	\partial_{\mu_q}\langle \hat \varphi_3 \rangle = &\frac{1}{\det V^{(2)}_\varphi}\left[ V_{\varphi, 38}  V_{\varphi,\mu_q 8} - V_{\varphi,88}  V_{\varphi,\mu_q 3} \right]\,,   \\[1ex] 
	\partial_{\mu_q} \langle \hat \varphi_8\rangle = & \frac{1}{\det V^{(2)}_\varphi}\left[ V_{\varphi, 38} V_{\varphi,\mu_q 3} - V_{\varphi,33} V_{\varphi,\mu_q 8}   \right] \,, 
	\label{eq:mu-depvarphi}
\end{align} 	
where the right hand side is evaluated at $\varphi_i = \langle \hat \varphi_i\rangle$. In \labelcref{eq:mu-depvarphi} we have used the short hand notations 
\begin{align} 
V_{\varphi,ij}=	\frac{\partial V_\varphi}{\partial \varphi_i \partial \varphi_j}\,, \qquad V_{\varphi,\mu_q i} = \frac{\partial V_\varphi}{\partial \mu_q  \partial \varphi_i}\,.
\end{align}
This concludes the discussion of the numerical computation of the Polyakov loop potential. Inserting the 
purely imaginary expectation value for $\varphi_8$ and the real one for $\varphi_3$,   
\begin{align} 
 \bigl(\bar \varphi_3\,,\, \imag 	\bar \varphi_8\bigr)=  \bigl( \langle \hat \varphi_3\rangle \,,\,\langle \hat \varphi_8\rangle\bigr) \,,\qquad \bar \varphi_3 \,,\, \bar\varphi_8 \in \mathbbm{R}\,,
\label{eq:varphi8muq}
 \end{align}
in \labelcref{eq:PbarPLoopAlg}, we arrive at 
\begin{align}\nonumber 
	{L}(\bar \varphi_3, \imag \bar \varphi_8) =&\, \frac{1}{3}\left[ e^{2 \pi \bar{\varphi}_8/\sqrt{3}} + 2 \, e^{- \pi \bar{\varphi}_8/\sqrt{3}} \cos{\pi\bar \varphi_3}\right]\,, \\[1ex] 
	\bar L(\bar \varphi_3,\imag\bar \varphi_8)  = &\,\frac{1}{3} \left[e^{- 2 \pi \bar{\varphi}_8/\sqrt{3}} + 2 \, e^{\pi \bar{\varphi}_8/\sqrt{3}} \cos{\pi\bar \varphi_3}\right]\,. 
\label{eq:LbarL-muq}
\end{align}
The order parameters $L,\bar L\in\mathbbm{R}$ in \labelcref{eq:LbarL-muq}  take real values. 
We close this discussion with the remark, that $L(\langle \hat \nu\rangle ),\bar L(\langle \hat \nu\rangle )$ have no direct physical meaning even in pure Yang-Mills theory. 
They serve a twofold purpose: they constitute optimised order parameters for center-symmetry breaking as well as encoding the optimal expansion point or gluonic background for the vertex expansion in functional approaches.  As such, they encode relevant dynamics of observables such as the thermodynamic observables discussed here. 

\begin{figure}[t]
	\centering
	\includegraphics[width=0.9\columnwidth]{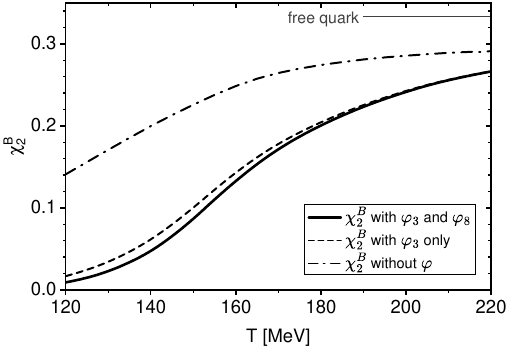}\\[2ex]
	\includegraphics[width=0.9\columnwidth]{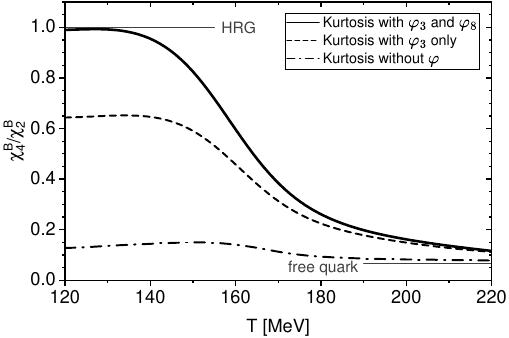}
	\caption{The impact of $\varphi^{3}$ and $\varphi^{8}$ background field on the baryon number susceptibilities $\chi_2^{B}$ and the kurtosis $\chi_4^{B}/\chi_2^{B}$ as a function of temperature at $\mu_B = 0$ and $\mu_Q = \mu_S = 0$. }
	\label{fig:chiB-A0-comp}
\end{figure}
% 

%%%%%%%%%%%%%%%%%%%%%%%%%%%%%%%%%%%%
\section{Further results on QCD thermodynamic functions}
\label{app:EoS}

Here we summarise some additional results on QCD thermodynamic functions which compare the case with and without $A_0$ feedback. 

First, we provide further details concerning the relevance of the dynamical background for the fluctuations of conserved charges, complementing the respective discussion in \Cref{sec:baryon-flucs}, see \Cref{fig:R42-comp-main}. In \Cref{fig:chiB-A0-comp}, we show the a more detailed comparison of
the second-order susceptibilities $\chi_2^{B}$ and the kurtosis $\chi_4^{B}/\chi_2^{B}$ with and without gluonic background. We compare the full computation with $\varphi^c$ with that with $\varphi^c=0$ already shown in \Cref{fig:R42-comp-main} as well as that with $\varphi^8=0$ and $\varphi^3\neq 0$: This background is obtained by solving the EoM for $\varphi^3$ at fixed $\varphi^8=0$ instead of solving the full equations of motion \labelcref{eq:EoMVvarphi}. We draw the following conclusions:
Both, $\varphi^3$ and $\varphi^8$, suppress the value of the baryon number susceptibilities at low temperatures, which are also responsible for recovering the low-temperature limit of the kurtosis and the limit $\chi_4^{B}/\chi_2^{B} = 1$ for weakly interacting baryons at $T=0$ is not obtained. 
As expected, the leading contribution for the $T=0$ limit comes from $\varphi^3$, while $\varphi^8$ gives a sub-leading contribution compared with $A_0^3$. Moreover, its effect is rather small in $\chi_2^{B}$, but it is not negligible in the kurtosis at low temperatures, see \Cref{fig:chiB-A0-comp}. 
We expect that the inclusion of the full dynamical background has an increasing importance for  higher-order fluctuations. 
We also expect that the gluonic background is of increasing importance for the QCD thermodynamics at larger baryon chemical potentials. 

\begin{figure}[t!]
	\centering
	\includegraphics[width=0.9\columnwidth]{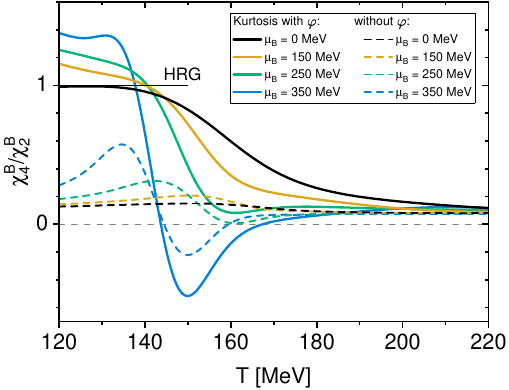}
	\caption{Baryon number kurtosis $\chi_4^{B}/\chi_2^{B}$ for $\mu_Q = \mu_S = 0$ as a function of temperature, with~(solid) and without~(dashed) the inclusion of the gluonic background $\varphi$. The comparison is given at several $\mu_B$ values: 0, 150, 250 and 350\,MeV. We also show the HRG limit $\chi_4^{B}/\chi_2^{B} = 1$ at low temperatures.}
	\label{fig:chiB-A0-comp-finite}
\end{figure}
\begin{figure*}[t]
	\centering
	\includegraphics[width=0.9\columnwidth]{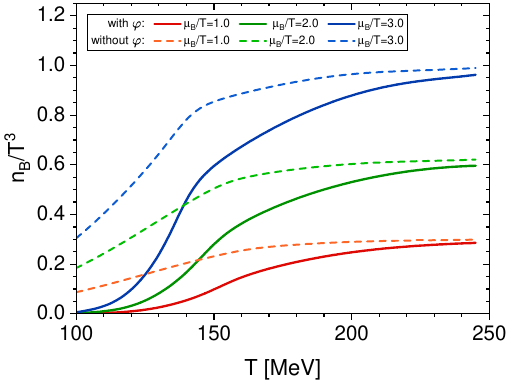}\hspace{.3cm}
	\includegraphics[width=0.865\columnwidth]{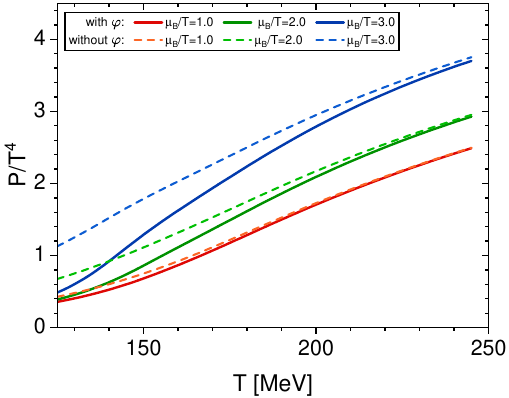}\\[3ex]
	\includegraphics[width=0.9\columnwidth]{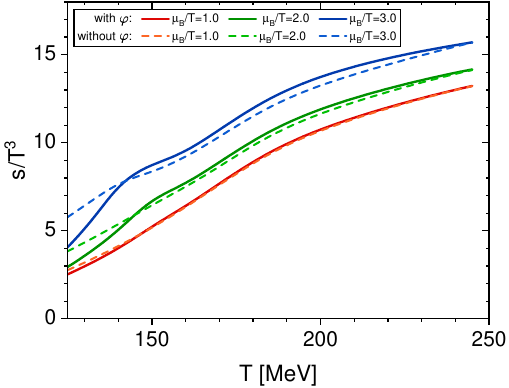}\hspace{.3cm}
	\includegraphics[width=0.9\columnwidth]{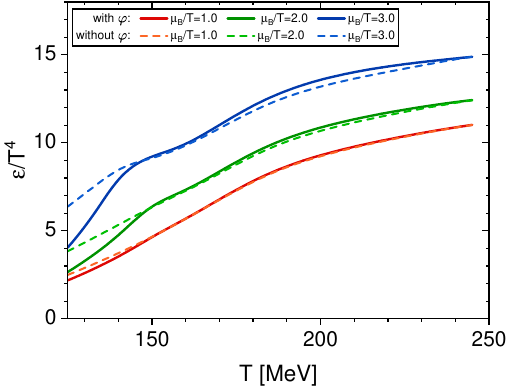}
	\caption{Impact of dynamical gluonic background on thermodynamic observables in QCD: baryon number density $n_B/T^3$, pressure $P/T^4$, entropy density $s/T^3$ and energy density $\epsilon/T^4$, as functions of temperature evaluated at $\mu_B/T=1.0,2.0,3.0$. }
	\label{fig:chiB-thermo-comp}
\end{figure*}

The respective analysis is summarised in \Cref{fig:chiB-A0-comp-finite} and \Cref{fig:chiB-thermo-comp} (thermodynamics): \\[-1ex] 

We first discuss the kurtosis shown in \Cref{fig:chiB-A0-comp-finite}: We conclude from our results that the kurtosis without the dynamical background fails to even show the qualitative features in the hadronic phase and its amplitude is off in the crossover regime and the hadronic regime. However, basic features such as the occurrence of non-monotonicity persist and the respective temperature regimes are comparable, even though only qualitatively.  

Finally, we analyse the relevance of the dynamical background for the kurtosis with a similar one for the net-baryon number density, pressure, entropy density and energy density discussed in \Cref{sec:nB+Thermodynamics}. This comparison is shown in \Cref{fig:chiB-thermo-comp} for $\mu_B/T=1.0,2.0, 3.0$. It is clear that the differences in the thermodynamic functions increase towards lower temperatures: the differences are sizeable for 
temperatures about and below the crossover temperature and grow larger for increasing baryon chemical potential. 

In summary, the inclusion of the gluonic background is of crucial importance for the qualitative behaviour of thermodynamic observables at finite baryon chemical potential. Moreover, its importance grows even stronger with increasing baryon chemical potential.

%%%%%%%%%%%%%%%%%%%%%%%%%%%%%%%%%%%%
\section{Computational Details}
\label{app:ComputationalDetails}

The computations of thermal sums can be schematically written as the Matsubara sum and the spatial-momentum integral of a specific kernel function $f$, 
\begin{align}
  \mathcal{F} = \sumint_{p} f(x),  \qquad  \sumint_{p} = T \sum_{\omega_{p,n}^{}}\int \frac{d^3 p}{(2\pi)^3}\,,
\end{align}
with $x=p^2$ the Laplacian. For computational convenience we consider 
the difference of the thermal sum~\cite{Gao:2020qsj,Gao:2020fbl}, 
\begin{align} \nonumber 
f(x;Z_{\phi}(x)) =&\, f(x;Z_{\phi}^{(0)})+  \Delta f\,,\\[1ex]
\Delta f =&\, f(x;Z_{\phi}(x)) - f(x;Z_{\phi}^{(0)})\,,
\end{align}
with a finite momentum cutoff $\Lambda$ and $Z_{\phi}^{(0)}$ is the wave function of the field $\phi$ at the cutoff. The thermal sum in  $f_{\phi}(x;Z_{\phi}^{(0)})$ can be evaluated analytically in the limit $\Lambda\to\infty$, 
\begin{align}
  \sumint_{p} f(x) =  \sumint_{p}{}^{\prime} \Delta f(x;Z_{\phi}(x)) + \sumint_{p} f(x;Z_{\phi}^{(0)})\,. 
  \label{app:standard-diff}
\end{align}
We take the $O(4)$-cutoff for the difference sum, 
\begin{equation}
  \sumint_{p}{}^{\prime} \quad \mbox{for} \quad \omega_{p,n}^2 + \spatial{p}^2 \leq \Lambda^2,
\end{equation}
and set its cutoff scale to be the same as the renormalisation scale of the propagators, which is $\Lambda = 40\,$GeV. 
For the sum of quark number densities \labelcref{eq:nq} at finite $\mu_q$ we apply an additional regularisation  for the spatial-momentum integral with the contour integration technique, see~\cite{Gao:2015kea, Isserstedt:2019pgx} for the details in Euclidean space.

It also turns out to be convenient to apply a further approximation to the quark mass function for achieving a better convergence on the numerical sums of \labelcref{eq:nq} at low temperatures, if combined with the contour techniques. 
For simplicity we illustrate this approximation in the case of vanishing background $\varphi$, that is $\bar A_0=0$. 
Without loss of generality, one can define the mass components $M_{q}^{+}$ and $M_{q}^{-}$ with the momentum pairs $p_{q}^{\pm} = (\pm\omega_{q,n} - \imag \mu_q, \spatial{p})$ where $\omega_{q,n} > 0$, 
\begin{align}
  M_{q,+}(p_{q}^{+}) = \frac{1}{2} \left[ M_{q}(p_{q}^{+})+M_{q}(p_{q}^{-}) \right]\,,  \\[1ex]
  M_{q,-}(p_{q}^{+}) = \frac{1}{2} \left[ M_{q}(p_{q}^{+})-M_{q}(p_{q}^{-}) \right]\,. \label{app:Mqplusminus}
\end{align}
The function $M_{q,\pm}$ are readily obtained from the mass function $M_q$. 
Conversely, $M_q$ can be re-expressed using $M_{q,+}$ and $M_{q,-}$ with the positive~(+) frequencies in $p_{q}$. The additional approximation concerns the $M_{q,-}$ component: an analytic expansion on $M_q$ with respect to the Laplacian $x_q = p_q^2$ in \labelcref{app:Mqplusminus} yields, 
\begin{align}
  M_{q,-} \approx - 2 \, \imag \, \omega_{q,n}\mu_q \frac{\partial M_q}{\partial x_q}, \label{app:Mqminus-expand}
\end{align}
which vanishes under the limit of $\omega_{q,n} \to 0$. However, in numerical calculations such an imaginary contribution from \labelcref{app:Mqminus-expand} results in a slow convergence of the spatial-momentum integral in \labelcref{eq:nq} at low $T$.
Hence, we neglect the $M_{q,-}$ component for the lowest frequencies $\omega_{q} = \pm \pi T$ and use instead, 
\begin{align}
  M_{q}(p_{q}^{\pm}) \approx M_{q,+}(p_{q}^{+}) \qquad   p_{q,0}^{\pm} = \pm \pi T - \imag \mu_q,  
  \label{app:Mqapprox}
\end{align}
in the kernel function $f_q(p_q,\mu_q; M_q)$ for the thermal sums of quark number densities. We have thoroughly tested the accuracy of this approximation and the respective error is less than 5\%. Finally, it can be used readily in the presence of the dynamical gluonic background: one simply replaces $\imag \mu_q$ with $\imag \mu_q - \, 2\pi \varphi^c$ in the above derivations. 

\hfill 
\newpage

\bibliography{A0_refs}

\end{document}